\documentclass[sigconf, nonacm]{acmart}
\newcommand\artifactavailabilityurl{https://anonymous.4open.science/r/DeriveMem-67C6}
\newcommand\evalframeworkurl{https://anonymous.4open.science/r/DerivEval-6799}

\usepackage{amsmath, amsthm}
\usepackage{multirow}
\usepackage{graphicx}
\usepackage[table]{xcolor}
\usepackage{siunitx}
\usepackage{enumitem}
\usepackage[linesnumbered,ruled,vlined]{algorithm2e}
\setlist[itemize]{leftmargin=1.2em, topsep=2pt, itemsep=1pt}
\setlist[enumerate]{leftmargin=1.2em, topsep=2pt, itemsep=1pt}
\usepackage{xspace}
\usepackage{pifont}
\usepackage{tcolorbox}
\usepackage{tikz}
\usetikzlibrary{arrows.meta,positioning,calc,fit,shapes.geometric,decorations.pathreplacing}
\usepackage{microtype}
\usepackage{booktabs}
\usepackage{flafter}

\definecolor{dmblue}{HTML}{0072B2}
\definecolor{dmorange}{HTML}{D55E00}
\definecolor{dmgreen}{HTML}{009E73}
\definecolor{dmdark}{HTML}{263238}
\definecolor{dmgray}{HTML}{66727A}
\definecolor{dmlightblue}{HTML}{EAF4FA}
\definecolor{dmlightorange}{HTML}{FCEFE9}
\definecolor{dmlightgreen}{HTML}{EAF6F2}
\definecolor{dmlightgray}{HTML}{F3F5F6}
\definecolor{dmsky}{HTML}{56B4E9}
\definecolor{dmamber}{HTML}{E69F00}
\definecolor{dmpurple}{HTML}{CC79A7}
\definecolor{fignavy}{HTML}{2F4F7F}
\definecolor{figochre}{HTML}{A0761C}
\definecolor{figgreen}{HTML}{2E6F57}

\tcbset{
  dmsemantic/.style={
    colback=dmlightblue,
    colframe=dmblue,
    boxrule=0.55pt,
    arc=1pt,
    left=4pt,
    right=4pt,
    top=3pt,
    bottom=3pt,
    before skip=5pt,
    after skip=5pt,
    fonttitle=\bfseries,
    coltitle=white
  },
  dmnote/.style={
    colback=dmlightgray,
    colframe=dmgray,
    boxrule=0.45pt,
    arc=1pt,
    left=4pt,
    right=4pt,
    top=3pt,
    bottom=3pt,
    before skip=4pt,
    after skip=4pt
  }
}

\theoremstyle{definition}
\newtheorem{definition}{Definition}

\newtheorem{example}{Example}
\newtheorem{property}{Property}

\newcommand{\sys}{\textsf{MemDerive}\xspace}          %
\newcommand{\sutname}[1]{{\fontfamily{lmtt}\selectfont #1}}
\newcommand{\memzero}{\sutname{Mem0}\xspace}
\newcommand{\cognee}{\sutname{Cognee}\xspace}
\newcommand{\lightmem}{\sutname{LightMem}\xspace}
\newcommand{\memoryos}{\sutname{MemoryOS}\xspace}
\newcommand{\bmtf}{\sutname{BM25}\xspace}
\newcommand{\bmtfE}{\sutname{BM25-E}\xspace}
\newcommand{\embed}{\sutname{Embed}\xspace}
\newcommand{\embedE}{\sutname{Embed-E}\xspace}
\newcommand{\zep}{\sutname{Zep}\xspace}
\newcommand{\locomo}{\textsf{LoCoMo}\xspace}
\newcommand{\longmemeval}{\textsf{LongMemEval}\xspace}

\newcommand{\tripops}{\texttt{TripOps}\xspace}
\newcommand{\shopops}{\texttt{ShopOps}\xspace}
\newcommand{\eduops}{\texttt{EduOps}\xspace}

\newcommand{\closedbook}{\sutname{Closed-Book}\xspace}
\newcommand{\fullmode}{\sutname{Full-Context}\xspace}
\newcommand{\oraclemode}{\sutname{Oracle}\xspace}

\newcommand{\kw}[1]{{\ensuremath{\mathsf{#1}}}\xspace}

\newcommand{\lblans}{\kw{answerable}}
\newcommand{\lblzero}{\kw{justified\text{-}zero}}
\newcommand{\lblunans}{\kw{unknown}}

\newcommand{\ansacc}{\kw{AnsAcc}}
\newcommand{\lblacc}{\kw{AnsblAcc}}
\newcommand{\witacc}{\kw{WitAcc}}
\newcommand{\stale}{\kw{StaleRate}}

\newcommand{\stitle}[1]{\par\smallskip\noindent{\bf #1}}
\newcommand{\resultfinding}[2]{%
  \begin{tcolorbox}[
    colback=gray!5,
    colframe=black,
    boxrule=0.5pt,
    arc=0.8pt,
    left=3pt,
    right=3pt,
    top=2.5pt,
    bottom=2.5pt,
    before skip=6pt,
    after skip=6pt]
  \small\textbf{Finding #1.} #2
  \end{tcolorbox}}
\newcommand{\dnum}[1]{\textcolor{fignavy}{\ding{#1}}}
\newcommand{\dstep}[2]{\dnum{#1}\hspace{0.25em}\textbf{#2}}
\newcommand{\yes}{\ding{51}}

\newcommand{\memgpt}{\sutname{MemGPT}\xspace}
\newcommand{\answerer}{\textsf{GPT-5.6 Sol}\xspace}  %

\newcommand{\ulabel}[1]{\underline{\smash{\textit{#1}}\rule[-0.6pt]{0pt}{0pt}}}

\newcommand{\NumWorlds}{10\xspace}
\newcommand{\NumQueriesTotal}{1{,}503\xspace}
\newcommand{\NumEventsTotal}{4{,}692\xspace}

\newcommand{\EOneBMAns}{58.28\xspace}
\newcommand{\EOneFullAns}{79.84\xspace}
\newcommand{\EOneOracleAns}{96.01\xspace}
\newcommand{\EOneMemoryAnsBand}{30.41--48.30\xspace}

\newcommand{\NumContextQueries}{400\xspace}
\newcommand{\EFourMemEight}{77.25\xspace}
\newcommand{\EFourBMHigh}{78.75\xspace}

\newcommand{\EFourFull}{80.00\xspace}
\newcommand{\EFourFullMedium}{93.25\xspace}

\newcommand{\EEightLMEMem}{84.00\xspace}
\newcommand{\EEightLMEBM}{78.60\xspace}
\newcommand{\EEightLoCoMoCognee}{68.13\xspace}
\newcommand{\EEightLoCoMoBM}{70.64\xspace}

\newcommand{\EThirteenSourcesOneBand}{73.19--93.48\xspace}
\newcommand{\EThirteenSourcesManyBand}{14.77--26.85\xspace}

\newcommand{\EThirteenMemNear}{83.42\xspace}
\newcommand{\EThirteenMemFar}{25.82\xspace}
\newcommand{\EThirteenMemFarStale}{20.33\xspace}

\newcommand{\ShapeMemZeroLow}{83.05\xspace}
\newcommand{\ShapeMemZeroHigh}{21.97\xspace}
\newcommand{\ShapeOracleHighWide}{86.80\xspace}
\newcommand{\ShapeMemoryHighWideBand}{15.25--25.22\xspace}

\newcommand{\ReplayBMPost}{80.00\xspace}
\newcommand{\ReplayLightMemPost}{47.50\xspace}
\newcommand{\ReplayMemoryMaxGain}{30.0\xspace}

\newcommand{\DiagUngatheredMemoryBand}{61--85\xspace}
\newcommand{\DiagUnretrievableLightMem}{83\xspace}
\newcommand{\DiagMisderivedBM}{38\xspace}
\newcommand{\DiagSelfRecoveryMemoryBand}{33--52\xspace}
\newcommand{\DiagSelfRecoveryLightMem}{9\xspace}
\newcommand{\DiagGoldRecoveryMemoryBand}{39--65\xspace}
\newcommand{\DiagGoldGainLightMem}{39\xspace}
\newcommand{\DiagRuleMemZero}{17\xspace}
\newcommand{\DiagRuleLightMem}{8\xspace}
\newcommand{\DiagDerivedBand}{26--61\xspace}
\newcommand{\DiagFactLightMem}{56\xspace}
\newcommand{\DiagTargetedBand}{86--94\xspace}
\newcommand{\DiagInheritedBand}{54--74\xspace}

\newcommand{\FirstCorrectDepthBM}{1.38\xspace}
\newcommand{\FirstCorrectDepthLightMem}{2.52\xspace}

\newcommand{\NumTrajectoryQueries}{132\xspace}
\newcommand{\TrajectoryBMFinal}{43.18\xspace}
\newcommand{\TrajectoryOracleFinal}{93.18\xspace}
\newcommand{\TrajectoryMemoryFirstBand}{28.03--65.15\xspace}
\newcommand{\TrajectoryMemoryFinalBand}{16.67--41.67\xspace}

\newcommand{\PreservationBand}{91.03--92.86\xspace}
\newcommand{\RefreshBand}{35.71--63.64\xspace}
\newcommand{\NumChangedPairs}{106\xspace}
\newcommand{\OracleUnrecovered}{1\xspace}
\newcommand{\MemoryUnrecoveredBand}{66--79\xspace}

\newcommand{\MemoryStaleBand}{9.43--18.87\xspace}

\newcommand{\CheckpointMemoryOSCost}{2.60\xspace}
\newcommand{\CheckpointBMLatency}{8.1\xspace}

\newcommand{\EndReadBMCost}{73.55\xspace}
\newcommand{\EndMemoryIngestCostBand}{0.31--2.57\xspace}
\newcommand{\EndMemoryReadCostBand}{9.13--37.67\xspace}

\newcommand{\MotMemNative}{86.7\xspace}
\newcommand{\MotMemDerived}{33.3\xspace}
\newcommand{\MotCogneeNative}{60.0\xspace}
\newcommand{\MotCogneeDerived}{16.7\xspace}
\newcommand{\MotFullDerived}{86.7\xspace}
\newcommand{\MotSubAnswerMax}{46.7\xspace}  %

\newcommand{\NumDomains}{3\xspace}
\newcommand{\ExampleWitnessSize}{44\xspace}
\newcommand{\NumStateProbes}{2{,}818\xspace}
\newcommand{\NumProbeRoots}{1{,}029\xspace}
\newcommand{\NumExplicitChanges}{18\xspace}
\newcommand{\NumImplicitChanges}{88\xspace}
\newcommand{\RefreshExplicitMemory}{82.4\xspace}
\newcommand{\RefreshImplicitMemory}{41.1\xspace}
\newcommand{\RefreshExplicitRetrieval}{83.3\xspace}
\newcommand{\RefreshImplicitRetrieval}{63.1\xspace}

\begin{document}

\title{\sys: A Diagnostic Benchmark for Agent Memory over Evolving Data}

\author{Sen Gao}
\affiliation{\institution{National University of Singapore}\city{Singapore}\country{Singapore}}
\email{sen@u.nus.edu}
\author{Yongan Li}
\affiliation{\institution{Shanghai Jiao Tong University}\city{Shanghai}\country{China}}
\email{liyongan@sjtu.edu.cn}
\author{Han Chen}
\affiliation{\institution{National University of Singapore}\city{Singapore}\country{Singapore}}
\email{chenhan@u.nus.edu}
\author{Bingsheng He}
\affiliation{\institution{National University of Singapore}\city{Singapore}\country{Singapore}}
\email{dcsheb@nus.edu.sg}
\author{Ming Wu}
\affiliation{\institution{Zero Gravity Labs}\city{Singapore}\country{Singapore}}
\email{ming@0g.ai}
\author{Jason Zeng}
\affiliation{\institution{Zero Gravity Labs}\city{Singapore}\country{Singapore}}
\email{jason@0g.ai}
\author{Michael Heinrich}
\affiliation{\institution{Zero Gravity Labs}\city{Singapore}\country{Singapore}}
\email{michael@0g.ai}
\author{Shixuan Sun}
\affiliation{\institution{Shanghai Jiao Tong University}\city{Shanghai}\country{China}}
\email{sunshixuan@sjtu.edu.cn}

\begin{abstract}
Long-running agents rely on memory of their past interactions for task state, such as what has been
completed and which conditions still hold. Such state is usually a derived
state. No record states it, and it must be computed by rules from multiple
records of an evolving history. Existing memory benchmarks score only the final answer,
with no reference for the intermediate states behind it. A wrong answer
therefore cannot be traced to the state that was missed, and a state that
silently became outdated goes undetected. We present \sys, a diagnostic benchmark for agent memory
over evolving data. It specifies each scenario symbolically, derives exact
gold answers with Datalog, records the gold derivation DAG and the supporting
events of every query, and renders the history in natural language with
fictional entities and distractors. Its \NumWorlds scenarios contain
\NumEventsTotal events and \NumQueriesTotal queries whose difficulty varies
along access, evolution, and composition. On this ground truth, \sys
evaluates answer reliability, diagnoses failures by probing the nodes of the
gold derivation DAG, and tracks whether answers stay correct as events arrive.

Our evaluation of four memory systems reveals failures that final-answer
benchmarks hide. On derived state, every memory system falls behind simple
\bmtf retrieval, although the same answer model reaches \EOneOracleAns\% when
given the supporting events. Most failures are ungathered, as a system holds
the missing intermediate state but does not retrieve it for the question that
needs it. Memory systems also rarely propagate implicit changes, which no new
record states. These results suggest three principles for memory design.
Gather inputs along the derivation, store rules with their versions, and
maintain derived states through recorded dependencies.
\end{abstract}

\maketitle
\raggedbottom
\pagestyle{plain}

\section{Introduction}\label{sec:intro}

\begin{figure}[t]
\centering
\includegraphics[width=\columnwidth]{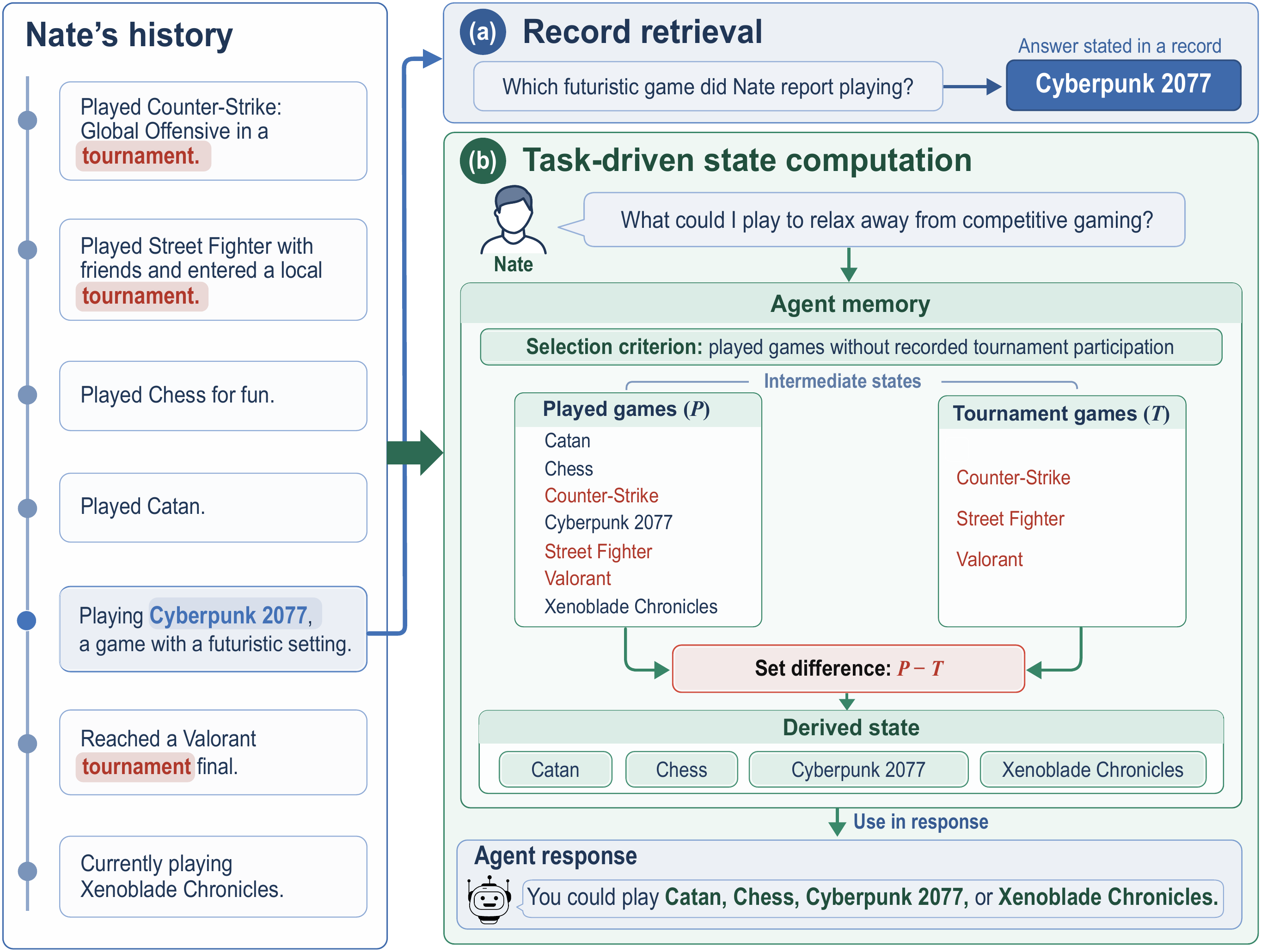}
\Description{Nate asks for games to play away from competitive gaming.
Agent memory computes the difference between played games and games with
recorded tournament participation, supplying four candidates for a response.}
\caption{An original question and a state question over the same LoCoMo
history. (a) One record states the answer. (b) The answer is computed from two
states, each gathered from several records of the history.}
\label{fig:motivation-pipeline}
\end{figure}

Long-running agents rely on memory of their past interactions to decide what
to do next~\cite{wang2023voyager,sumers2024coala}. A decision
depends on what has been completed, which conditions still hold, and what
remains to be done. Anthropic reports coding agents that prematurely consider
a project complete and addresses this with persistent progress records and
verification across sessions~\cite{young2025harnesses}. Generative
agents~\cite{park2023generativeagents} and Reflexion~\cite{shinn2023reflexion}
likewise synthesize past observations into reflections that guide later
behavior. In both cases the agent needs a task
state that the records of its history support but rarely state. Memory
systems such as \memzero~\cite{chhikara2025mem0} and
\cognee~\cite{markovic2025cognee} extract information from the history and
retrieve what matches a request.
This paper asks whether they supply the state that a task needs.

\textls[-3]{Figure~\ref{fig:motivation-pipeline} shows the difference on a LoCoMo history.
The original question asks which futuristic game Nate reported playing, and
one record answers it. Now Nate asks what he could play to relax away from
competitive gaming. A good answer lists the games he has played without any
recorded tournament. No record states this set. Memory must first build two
intermediate states from several sessions, the games Nate has played and the
games he played in tournaments, and then take their difference. We call a
fact that no record states but that follows from several records by rules a
\emph{derived state}, and a question whose answer is a derived state a
\emph{state question}. The status of a task is a typical derived state. Like
the result of a view over database tables, a derived state is computed from the
history rather than stored in it. Here, however, the base data is natural
language, and the view is defined only when a question arrives.}

Derived state is hard for memory systems. The records that support it
rarely share words with the question and have low embedding similarity to it.
They matter only in combination, and together they can outnumber any
retrieval budget. A derived state is usually composed from intermediate
states layer by layer, so one wrong intermediate state corrupts every state
above it. It can also become stale without any record saying so. A later
record that Nate entered a Chess tournament would remove Chess from the answer,
although it never mentions the games he could play to relax. These properties
affect every way of building memory, whether a system retrieves raw records
for each question, extracts and stores facts as events arrive, or searches
and reasons over the history when a question arrives.

\begin{figure}[t]
\centering
\includegraphics[width=\columnwidth]{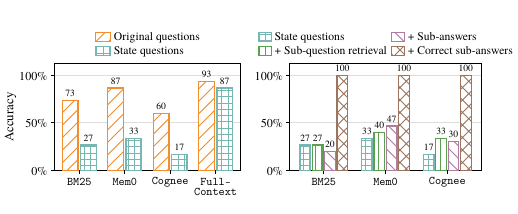}\par
\vspace{-6pt}
\makebox[0.52\columnwidth]{\small\textbf{(a) Accuracy by question type}}%
\makebox[0.48\columnwidth]{\small\textbf{(b) Answering via sub-questions}}
\Description{Panel A shows accuracy on original LoCoMo and LongMemEval
questions and on added state questions over the same histories. Panel B
splits each state question into two sub-questions and shows accuracy when the
state question is answered directly, with retrieval per sub-question, from the
system's sub-answers, and from correct sub-answers.}
\caption{Accuracy on LoCoMo and LongMemEval histories. A state question asks
for an answer that no single record states and that must be computed from
several records.}
\label{fig:empirical-teaser}
\end{figure}

A paired study shows that current memory systems fail on state questions. We
take histories from LoCoMo and LongMemEval, keep an original question, and
add a state question over the same history and the same stored memories.
Figure~\ref{fig:empirical-teaser}(a) shows \memzero falling from
\MotMemNative\% to \MotMemDerived\% and \cognee from \MotCogneeNative\% to
\MotCogneeDerived\%. Given the whole history, the same answer model still
reaches \MotFullDerived\%, so the questions are answerable and the loss lies
in what memory supplies. Figure~\ref{fig:empirical-teaser}(b) splits each
state question into two sub-questions, one for each intermediate state.
Retrieving separately for each sub-question helps some systems, and so does
answering the state question from the system's own answers to the
sub-questions, but no system exceeds \MotSubAnswerMax\%. Given the correct
answers to the sub-questions, the answer model solves every state question.
The bottleneck is therefore obtaining the intermediate states, not combining
them, and an evaluation that scores only final answers cannot see it.

Existing benchmarks were not designed to expose this failure. LoCoMo and
LongMemEval test recall, temporal and multi-session reasoning, and knowledge
updates over long conversations~\cite{maharana2024locomo,wu2025longmemeval}.
StateMemBench separates current from superseded
answers~\cite{fan2026statemembench}, and MemTrace audits how evidence is
reached and used~\cite{long2026memtrace}. These benchmarks judge the final
answer. Their histories are written text without executable semantics, so
they cannot supply the correct value of an intermediate state, and a wrong
answer cannot be traced to the step that failed
(Table~\ref{tab:comparison}). Their updates are explicit, since a new
record revises a fact and states its new value. A derived state also changes
implicitly, when a fact it depends on changes and no record mentions the
state. Rankings on these benchmarks therefore do not carry over to derived
state. On LongMemEval, \memzero reaches \EEightLMEMem\% and exceeds \bmtf at
\EEightLMEBM\%, yet on derived state every memory system falls behind \bmtf.

\stitle{Our approach.}
We present \sys, a diagnostic benchmark for agent memory over evolving data.
Its design follows from four needs.
\ulabel{Exact ground truth.} Every intermediate state needs a correct value at
every point of an evolving history. \sys therefore specifies each scenario
symbolically. Each event pairs a natural-language record with symbolic
actions, and a Datalog program with aggregation derives every state from the
current facts and rules, so the gold is exact by construction.
\ulabel{Derivations as evidence.} For each query, \sys records a gold derivation
DAG that links the answer through its intermediate states to the events, and
the witnesses, the minimal sets of events that support the answer.
\ulabel{No shortcuts.} The answer must come from the history alone. No record
states a derived state, every recognizable entity is fictional so that
pretrained knowledge cannot help, and the key terms of a question often
appear only in distractor records.
\ulabel{Controlled difficulty.} Generation varies three workload dimensions
independently, namely Access to the support, Evolution of the facts, and
Composition of the derivation, so that a failure can be traced to one of
them. The suite covers travel reimbursement, shopping, and degree planning in
\NumWorlds scenarios with \NumEventsTotal events and \NumQueriesTotal queries.

Memory systems share no interface beyond answering questions, so \sys
observes them as black boxes through three evaluation questions.
\textbf{Q1: Reliability} compares answers exactly with the gold and citations
with the witnesses. \textbf{Q2: Diagnosis} turns the nodes of each gold
derivation DAG into state probes, questions about single intermediate states.
Following the DAG from the answer toward the events, it locates the lowest
states that the system gets wrong, classifies why they fail, and tests whether
repairing them recovers the answer. \textbf{Q3: Maintenance} asks tracked queries at successive
checkpoints and separates the refresh of changed answers from the
preservation of unchanged ones.

\stitle{Findings and implications.}
We evaluate four memory systems and two raw-retrieval baselines under one
answer model. First, memory systems answer \EOneMemoryAnsBand\% of the
queries, below \bmtf at \EOneBMAns\%, whereas the whole history yields
\EOneFullAns\% and the supporting events \EOneOracleAns\%. Accuracy falls along
all three workload dimensions, also under matched context budgets. Second,
most failures are \emph{ungathered}. The system answers a missing
intermediate state correctly when asked about it directly, but does not
retrieve it for the question that needs it. Rebuilding the answer from these
intermediate answers recovers up to \ReplayMemoryMaxGain points of accuracy,
so a memory system should split a request for a
derived state into requests for its inputs. Some systems fail to return rules
even when asked directly, so rules should be stored with their versions, like
facts. Third, memory systems preserve \PreservationBand\% of unchanged answers
and refresh \RefreshExplicitMemory\% of explicit changes, which a new record
states, but only \RefreshImplicitMemory\% of implicit changes, which propagate
through dependencies. Stored states therefore need recorded dependencies, as
in incremental view maintenance~\cite{gupta1993dred,motik2015bf}. Their work
at ingestion buys neither accuracy nor refresh.

Our contributions are as follows.
\begin{itemize}[leftmargin=1.4em]
\item \ulabel{Problem characterization.} We define derived state
over an evolving history, identify the properties that make it hard for
memory systems, and derive three workload dimensions, connecting agent memory
to views and provenance (Section~\ref{sec:background}).
\item \ulabel{Benchmark construction.} Executable scenarios
generate histories and queries with exact gold derivations, witnesses, and
controlled workload dimensions (Section~\ref{sec:construction}).
\item \ulabel{Evaluation methodology.} State probes in
two contexts locate and classify failures, two recoveries test them, and
tracked queries separate refresh from preservation
(Section~\ref{sec:methodology}).
\item \ulabel{Experimental study.} A comprehensive evaluation of four
memory systems and two retrieval baselines over a broad set of metrics yields
three findings and design lessons for memory systems
(Sections~\ref{sec:results} and~\ref{sec:lessons}).
\end{itemize}

\begin{sloppypar}
\stitle{Reproducibility.}
The benchmark dataset is available at \url{\artifactavailabilityurl}, and the
evaluation framework that runs all experiments is available at
\url{\evalframeworkurl}.
\end{sloppypar}

\section{Derived State over Evolving Data}\label{sec:background}\label{sec:problem}

This section defines derived state over an evolving history and derives its
workload dimensions and its implications for memory systems and their
evaluation. Table~\ref{tab:notation} summarizes the notation.

\begin{table}[t]
\caption{Notation and defining sections.}
\label{tab:notation}
\footnotesize\centering
\setlength{\tabcolsep}{3pt}
\begin{tabular}{@{}llc@{}}
\toprule
Symbol & Meaning & Sec.\\
\midrule
\(e=(t_e,r_e,A_e)\) & event: observation time, record, actions & \ref{sec:data-model}\\
\(O\), \(O_{\le\tau}\) & history, events observed up to cutoff \(\tau\) & \ref{sec:data-model}\\
\(B_\tau\) & resolved state, latest version of each fact and rule & \ref{sec:data-model}\\
\(P_\tau\), \(C_\tau\) & derivation program, closure & \ref{sec:closure}\\
\(q=(\theta,\tau)\) & query with target \(\theta\) and cutoff \(\tau\) & \ref{sec:closure}\\
\(z(q)=(\ell_q,y_q)\) & gold answer, label and answer & \ref{sec:closure}\\
\(\mathcal G_q\) & gold derivation DAG & \ref{sec:provenance}\\
\(S(q)\), \(\mathcal W(q)\) & support, witnesses & \ref{sec:provenance}\\
\(\hat E_q\) & events cited by the SUT & \ref{sec:reliability}\\
\(p_E\), \(r_W\) & citation precision, citation recall & \ref{sec:reliability}\\
\(\mathcal D_q\) & diagnostic DAG of state probes & \ref{sec:diagnosis}\\
\(i_s(u\mid v)\), \(t_s(v)\) & probe of \(u\) correct in the context of \(v\), and when targeted & \ref{sec:diagnosis}\\
\(\tau_j\), \(q_j\) & checkpoint \(j\), query instance at \(\tau_j\) & \ref{sec:maintenance}\\
\(c_{s,j}(q)\), \(d_j(q)\) & correct answer at \(\tau_j\), changed transition at \(\tau_j\) & \ref{sec:maintenance}\\
\bottomrule
\end{tabular}
\end{table}

\subsection{Data Model}\label{sec:data-model}\label{sec:benchmark-anatomy}

A memory system ingests a history of events.

\begin{definition}[Event and history]\label{def:event}
\textls[-2]{An \emph{event} \(e\) is a triple \((t_e,r_e,A_e)\) of an observation time
\(t_e\), a natural-language record \(r_e\), and a nonempty set \(A_e\) of
\emph{actions}. Each event has a unique identifier. An action asserts a fact
or declares a rule. A \emph{history} \(O\) is a set of events.}
\end{definition}

\noindent A memory system reads only the records \(r_e\) and cites events by
their identifiers. Gold answers depend only on the actions \(A_e\).

\stitle{Versions and cutoff.}
A fact can be asserted more than once. Each assertion creates a
\emph{version} of the fact, and a revision or cancellation supersedes the
previous version, which then becomes \emph{obsolete}~\cite{snodgrass1985taxonomy}. Which
versions are visible depends on time. Every query fixes an \emph{observation
cutoff} \(\tau\), and the events known at that point are
\(O_{\le\tau}=\{e\in O : t_e\le\tau\}\).

\begin{definition}[Resolved state]\label{def:resolved-state}
\textls[-5]{The \emph{resolved state} \(B_\tau=\kw{Resolve}(O_{\le\tau})\) contains, for
every fact, the latest version visible at \(\tau\):}
\begin{equation}\label{eq:resolved-state}
\kw{cur}(f,e) \leftarrow e\in O_{\le\tau},\ f\in A_e,\
\neg\,\kw{superseded}(f,e,\tau).
\end{equation}
\end{definition}

\noindent Here \(f\in A_e\) means that an action of \(e\) asserts a version of
the fact \(f\), and \(\kw{superseded}(f,e,\tau)\) holds when a later event in
\(O_{\le\tau}\) asserts a newer version of \(f\). Rules declared by events are
versioned and resolved in the same way, so \(B_\tau\) also holds the current
version of each rule.

\begin{example}[Vera's business trips]\label{ex:travel-state}
The history below is abridged from the travel domain of \sys, and all places
are fictional. Each event pairs a record, which an SUT reads, with the fact its
action asserts. The obsolete \(e_7\) is struck out, \(e_{10}\) supersedes it,
and the distractor \(e_{11}\) asserts nothing.
\par\smallskip
\noindent{\footnotesize\setlength{\tabcolsep}{2.5pt}\renewcommand{\arraystretch}{1.05}%
\newcommand{\ent}[1]{\textbf{#1}}%
\newcommand{\strike}[1]{\sbox0{#1}\rlap{\rule[0.55ex]{\wd0}{0.4pt}}#1}%
\begin{tabular}{@{}l@{\hspace{3pt}}>{\raggedright\arraybackslash}p{0.535\columnwidth}>{\raggedright\arraybackslash}p{0.305\columnwidth}@{}}
\toprule
Event & Record & Action\\
\midrule
\(e_1\) Jan & \ent{Ashfell} is a city in \ent{Corvia}. & \(\kw{cityIn}(\text{Ashfell},\text{Corvia})\)\\
\(e_2\) Jan & \ent{Corby} is a city in \ent{Corvia}. & \(\kw{cityIn}(\text{Corby},\text{Corvia})\)\\
\(e_3\) Jan & \ent{Lunmar} is in \ent{Meleth}. & \(\kw{cityIn}(\text{Lunmar},\text{Meleth})\)\\
\(e_4\) Jan & \ent{Halewick} is in \ent{Drossel}. & \(\kw{cityIn}(\text{Halewick},\!\text{Drossel})\)\\
\(e_5\) Jan & HR record: \ent{Vera} is based in \ent{Ashfell}. & \(\kw{basedIn}(\text{Vera},\text{Ashfell})\)\\
\(e_6\) Feb & Lunmar Harbor Inn booked for \ent{Trip-Anser}. & \(\kw{dest}(\text{Anser},\text{Lunmar})\)\\
\textcolor{dmgray}{\(e_7\) Mar} & \textcolor{dmgray}{\strike{Halewick Lodge booked for Trip-Grus.}} & \textcolor{dmgray}{\strike{\(\kw{dest}(\text{Grus},\text{Halewick})\)}}\\
\(e_8\) Apr & Halewick Lodge booked for \ent{Trip-Ibex}. & \(\kw{dest}(\text{Ibex},\text{Halewick})\)\\
\(e_9\) Jun & Vera's 2026 travel record lists \ent{Trip-Anser}, \ent{Trip-Grus}, \ent{Trip-Ibex}, and no other trips. & \(\kw{trip}(\text{Vera},\cdot)\) for the three, \(\kw{complete}(\kw{trip})\)\\
\textcolor{dmorange}{\(e_{10}\) Jul} & Halewick reservation for \ent{Trip-Grus} canceled. Corby Overlook Hotel booked instead. & \textcolor{dmorange}{\(\kw{dest}(\text{Grus},\text{Corby})\)}\\
\textcolor{dmgray}{\(e_{11}\) Aug} & \textcolor{dmgray}{\textit{Chat: Vera is abroad more than at her desk.}} & \textcolor{dmgray}{none}\\
\bottomrule
\end{tabular}}
\par\smallskip
\noindent The query asks \emph{how many of Vera's 2026 business trips go
outside her home country}. At an August cutoff, the destination of Trip-Grus
is Corby.
\end{example}

\subsection{Derived State and Queries}\label{sec:closure}

\begin{definition}[Derivation program, closure, derived state]\label{def:derived-state}
The \emph{derivation program} \(P_\tau\) consists of the fixed rules of the
domain and the current rules in \(B_\tau\). It is a Datalog program extended
with stratified negation and
aggregation~\cite{abiteboul1995foundations,apt1988declarative,consens1993aggregation}.
The \emph{closure} \(C_\tau=\kw{Derive}(P_\tau,B_\tau)\) is the set of facts
that \(P_\tau\) derives from \(B_\tau\), and a \emph{derived state} is a fact
in \(C_\tau\setminus B_\tau\).
\end{definition}

The closure is obtained by applying the rules to the resolved state until a
fixpoint is reached. It holds everything the history entails at \(\tau\) and
can be very large, so the benchmark never materializes it and computes only
the states that its queries need. Because events declare rules, \(P_\tau\)
evolves together with the data.

\begin{example}[Rules for Example~\ref{ex:travel-state}]\label{ex:travel-rules}
Three rules derive the target. Each rule derives its left side
whenever its right side holds, and \kw{dest} gives the current booking city
of a trip in \(B_\tau\):
{\small
\begin{align*}
\kw{country}(T,N) &\leftarrow \kw{dest}(T,C),\ \kw{cityIn}(C,N)\\
\kw{home}(P,N) &\leftarrow \kw{basedIn}(P,C),\ \kw{cityIn}(C,N)\\
\kw{outside}(P,T) &\leftarrow \kw{trip}(P,T),\ \kw{country}(T,N),\\
 &\phantom{\leftarrow{}}\ \kw{home}(P,M),\ N\neq M.
\end{align*}}%
The query counts the facts \(\kw{outside}(\text{Vera},T)\). In August the
count is two, for Trip-Anser and Trip-Ibex.
\end{example}

\begin{definition}[Query and gold answer]\label{def:task-question}\label{def:root-query}
A \emph{query} \(q=(\theta,\tau)\) is a natural-language question that asks
for a derived state \(\theta\), called its \emph{target}, at cutoff \(\tau\).
Its
\emph{gold answer} \(z(q)=(\ell_q,y_q)\) consists of an answerability label
\(\ell_q\) and an answer \(y_q\), both determined by \(C_\tau\). The answer can
be an entity, a set, a Boolean, or a number.
\end{definition}

\stitle{Answerability.}\label{sec:answerability}
Not every query can be answered from the history. The label \(\ell_q\) is
\lblans when the history determines the answer, \lblzero when it determines
that the answer is zero or empty, and \lblunans when it does not determine
the answer, in which case \(y_q=\bot\). For a count, total, or empty set, the
history determines the answer only if some event asserts that the collection
is complete~\cite{reiter1978closed,levy1996complete,razniewski2011completeness}.

\subsection{Gold Derivations and Witnesses}\label{sec:provenance}\label{sec:witnesses}

An answer alone cannot show which inputs a system used or where it went
wrong. \sys therefore records, for every query, how its gold answer follows
from the history.

\begin{definition}[Gold derivation DAG]\label{def:gold-proof-dag}
The \emph{gold derivation DAG} \(\mathcal G_q\) of query \(q\) is a finite
directed acyclic graph whose root is the target of \(q\). Its inner nodes are
the derived states on which the root depends, called \emph{intermediate
states}, together with the resolved facts and rules used to derive the root.
Its leaves are the events that assert these facts and declare these rules.
Each edge points from a node to an input it needs, called a \emph{child}.
When the answer is \lblunans, a leaf marks the missing fact instead of an
event.
\end{definition}

A shared input can serve several parents, so the structure is a DAG rather
than a tree. The \emph{height} of \(\mathcal G_q\) is the length of its longest path from
the root, and its \emph{width} is the largest number of nodes at the same
depth. Height measures how
deep a derivation is, and width measures how many inputs it combines at one
level. Because every intermediate state is a node with a known gold state, an
answer can be checked step by step rather than only at the root.

\begin{definition}[Support and witness]\label{def:witness}
The \emph{support} \(S(q)\) of query \(q\) is the set of events that contribute
to some derivation of \(z(q)\). A
\emph{witness} of \(q\) is a minimal subset \(W\subseteq S(q)\) from which
\(z(q)\) can be derived, and \(\mathcal W(q)\) denotes the set of all witnesses
of \(q\).
\end{definition}

A query can have several witnesses, because several events can assert the
same fact. The
definition of a witness follows minimal why-provenance~\cite{buneman2001why}.
An event outside \(S(q)\) whose record resembles relevant information is a
\emph{distractor} for \(q\).

\begin{figure}[t]
\centering
\resizebox{\columnwidth}{!}{%
\begin{tikzpicture}[font=\small,
 /utils/exec={%
   \definecolor{pink}{HTML}{222222}%
   \definecolor{pgray}{HTML}{8A8A8A}%
   \definecolor{pfill}{HTML}{EDEDED}%
   \definecolor{pcomp}{HTML}{2E6F57}%
   \definecolor{pcompfill}{HTML}{EAF2EE}%
   \definecolor{pevol}{HTML}{A0761C}%
   \definecolor{pevolfill}{HTML}{F7F0E1}%
   \definecolor{pacc}{HTML}{2F4F7F}%
   \definecolor{paccent}{HTML}{B23A2B}%
   \definecolor{paccentfill}{HTML}{F8E6E3}},
 node/.style={draw=pink,line width=0.55pt,rounded corners=1pt,align=center,
   inner xsep=2.5pt,inner ysep=1.6pt,font=\large,text=pink},
 dv/.style={node,draw=pcomp,fill=pcomp!17},
 rf/.style={node,draw=pevol,fill=pevol!20},
 ev/.style={circle,draw=pacc,fill=white,line width=0.55pt,inner sep=0pt,
   minimum size=0.66cm,font=\Large,text=pink},
 old/.style={ev,draw=pgray,text=pgray},
 noise/.style={old,densely dashed},
 fresh/.style={ev,draw=paccent,fill=paccentfill,text=paccent,line width=1pt},
 arr/.style={-{Latex[length=5pt,width=3.8pt]},draw=pink,line width=0.55pt},
 rule/.style={draw=pgray,line width=0.4pt,dash pattern=on 2pt off 1.5pt},
 layer/.style={rotate=90,font=\small\scshape\bfseries,anchor=center},
 key/.style={font=\large,anchor=west,text=pink}]
\fill[pcomp!8] (-7.47,0.6) rectangle (6.35,-2.92);
\fill[pevol!10] (-7.47,-2.92) rectangle (6.35,-4.62);
\fill[pacc!8] (-7.47,-4.62) rectangle (6.35,-5.98);
\draw[rule] (-7.47,-2.92) -- (6.35,-2.92);
\draw[rule] (-7.47,-4.62) -- (6.35,-4.62);
\draw[pcomp,line width=0.8pt] (-7.0,0.45) -- (-7.0,-2.8);
\draw[pevol,line width=0.8pt] (-7.0,-3.02) -- (-7.0,-4.52);
\draw[pacc,line width=0.8pt] (-7.0,-4.72) -- (-7.0,-5.88);
\node[layer,text=pcomp] at (-7.24,-1.2) {\ding{204}\ Composition};
\node[layer,text=pevol] at (-7.24,-3.77) {\ding{203}\ Evolution};
\node[layer,text=pacc] at (-7.24,-5.3) {\ding{202}\ Access};
\node[dv] (root) at (0,0) {\(\kw{count}=\mathbf{2}\)};
\node[dv] (out) at (0,-1.1) {\(\kw{outside}(\text{Vera})=\{\text{Anser},\text{Ibex}\}\)};
\node[dv] (ca) at (-1.85,-2.35) {\(\kw{country}\)\\(Anser)};
\node[dv] (cg) at (0.15,-2.35) {\(\kw{country}\)\\(Grus)};
\node[dv] (ci) at (3.35,-2.35) {\(\kw{country}\)\\(Ibex)};
\node[dv] (home) at (4.9,-2.35) {\(\kw{home}\)\\(Vera)};
\node[rf] (cmp) at (-6.0,-3.78) {\(\kw{complete}\)\\(\(\kw{trip}\))};
\node[rf] (trips) at (-3.75,-3.78) {\(\kw{trip}(\text{Vera})=\)\\\{Anser, Grus, Ibex\}};
\node[rf] (dg) at (-0.45,-3.78) {\(\kw{dest}\)\\(Grus, \textcolor{paccent}{\textbf{Corby}})};
\node[rf] (cc) at (1.95,-3.78) {\(\kw{cityIn}\)\\(Corby, Corvia)};
\node[rf] (bv) at (4.9,-3.78) {\(\kw{basedIn}\)\\(Vera, Ashfell)};
\node[ev] (e9) at (-4.4,-5.3) {\(e_9\)};
\node[ev] (e6) at (-2.35,-5.3) {\(e_6\)};
\node[ev] (e3) at (-1.65,-5.3) {\(e_3\)};
\node[old] (e7) at (-0.95,-5.3) {\(e_7\)};
\node[fresh] (e10) at (-0.2,-5.3) {\(e_{10}\)};
\node[ev] (e2) at (1.95,-5.3) {\(e_2\)};
\node[ev] (e8) at (3.2,-5.3) {\(e_8\)};
\node[ev] (e4) at (3.9,-5.3) {\(e_4\)};
\node[ev] (e5) at (4.6,-5.3) {\(e_5\)};
\node[ev] (e1) at (5.92,-5.3) {\(e_1\)};
\node[noise] (e11) at (5.26,-5.3) {\(e_{11}\)};
\draw[pgray,line width=0.5pt] (e7.north west) -- (e7.south east);
\draw[pgray,line width=0.5pt] (e7.north east) -- (e7.south west);
\draw[arr] (root) -- (out);
\draw[arr] (root.west) -| (cmp.north);
\draw[arr] (out.west) -| (trips.north);
\foreach \n in {ca,cg,ci,home} \draw[arr] (out) -- (\n.north);
\draw[arr] (cmp) -- (e9);
\draw[arr] (trips) -- (e9);
\draw[arr] (ca) -- (e6);
\draw[arr] (ca) -- (e3);
\draw[arr] (cg) -- (dg);
\draw[arr] (cg) -- (cc);
\draw[arr,draw=paccent,line width=0.8pt] (dg) -- (e10);
\draw[arr,densely dashed,draw=pgray] (dg) -- (e7);
\draw[arr] (cc) -- (e2);
\draw[arr] (ci) -- (e8);
\draw[arr] (ci) -- (e4);
\draw[arr] (home) -- (bv);
\draw[arr] (home.east) .. controls (6.32,-2.4) and (6.22,-4.4) .. (e1.north);
\draw[arr] (bv) -- (e5);
\node[font=\large,text=pink,anchor=west,align=left] at (1.0,0.05)
 {June \textbf{3} \(\to\) August \textcolor{paccent}{\textbf{2}},\\
  \textit{since \(e_{10}\) supersedes \(e_7\)}};
\draw[pgray,line width=0.4pt] (-7.47,-6.1) -- (6.35,-6.1);
\node[dv,minimum width=0.5cm,minimum height=0.32cm] at (-6.85,-6.42) {};
\node[key] at (-6.55,-6.42) {derived state};
\node[rf,minimum width=0.5cm,minimum height=0.32cm] at (-4.15,-6.42) {};
\node[key] at (-3.85,-6.42) {resolved fact};
\node[ev,minimum size=0.44cm] at (-1.45,-6.42) {};
\node[key] at (-1.18,-6.42) {event};
\node[old,minimum size=0.44cm] (k2) at (0.25,-6.42) {};
\draw[pgray,line width=0.5pt] (k2.north west) -- (k2.south east);
\draw[pgray,line width=0.5pt] (k2.north east) -- (k2.south west);
\node[key] at (0.52,-6.42) {obsolete};
\node[fresh,minimum size=0.44cm] at (2.25,-6.42) {};
\node[key] at (2.52,-6.42) {new};
\node[noise,minimum size=0.44cm] at (3.65,-6.42) {};
\node[key] at (3.92,-6.42) {distractor};
\end{tikzpicture}}
\Description{The gold derivation DAG of the running example in three bands.
Events at the bottom form the Access layer, resolved facts in the middle the
Evolution layer, and derived states at the top the Composition layer. The
count depends on the foreign-trip set and on the completeness of the trip
list. The obsolete event e7 is crossed out, e10 supersedes it in July and
changes the count from 3 to 2, and e11 is a distractor. Resolved facts for
Trip-Anser, Trip-Ibex, and Ashfell are drawn as their events.}
\caption{Gold derivation DAG of Example~\ref{ex:budget-derivation}, layered
by the three workload dimensions.}
\label{fig:proof-dag}\label{fig:dimensions}
\end{figure}
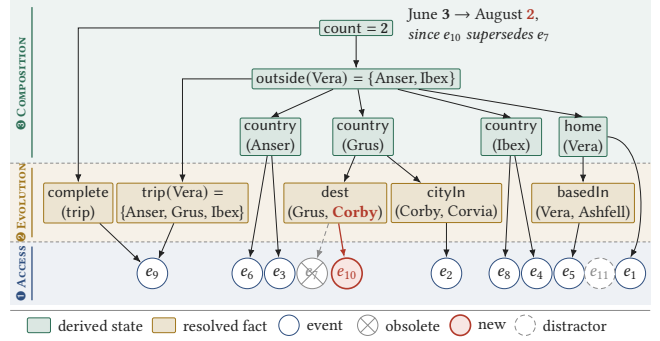

\begin{example}[Gold derivation for Example~\ref{ex:travel-state}]\label{ex:budget-derivation}
Figure~\ref{fig:proof-dag} shows \(\mathcal G_q\) at the August cutoff. The
only witness is \(\{e_1,\ldots,e_6,e_8,e_9,e_{10}\}\). It excludes the obsolete
\(e_7\) and the distractor \(e_{11}\) but still contains nine of the eleven
events. Every fact in this history is asserted by a single event, so \(S(q)\)
equals the only witness. Before \(e_9\) arrives, no event asserts that the trip
list is complete, so the label is \lblunans, although three foreign trips are
already visible.
\end{example}

\subsection{Characteristics and Workload Dimensions}\label{sec:dimensions}\label{sec:workload-dimensions}\label{sec:properties}

A recorded fact is stated by one record, rests on that record alone, depends
on no other state, and changes only when an event asserts a new version of
it. Derived state differs in all four respects.

\begin{property}[Implicitness]\label{prop:implicit}
No record states a derived state, so it can only be computed from facts by
rules. A query uses the vocabulary of derived predicates, whereas records use
the vocabulary of recorded facts.
\end{property}

\begin{property}[Joint support]\label{prop:support}
A derived state rests on a set of events that are relevant only together. The
set can be large, and the answer needs all of it.
\end{property}

\begin{property}[Layered dependence]\label{prop:exact}
A derived state is composed from intermediate states, layer by layer, and the
same intermediate state is shared by many queries. If a system derives an
intermediate state incorrectly, the error can propagate to every state above
it.
\end{property}

\begin{property}[Indirect change]\label{prop:nonmono}
A derived state changes whenever one of its inputs, a fact or a rule, changes,
even if no new event mentions it. One event can therefore change many derived
states at once.
\end{property}

\begin{example}[Characteristics of Example~\ref{ex:travel-state}]\label{ex:travel-difficulty}
All four characteristics appear in the running example. No record states the
count. The full \tripops query behind the example asks about trips
\emph{abroad}, and the only records containing that word are distractors such
as \(e_{11}\). The Corby booking
in \(e_{10}\) is relevant only together with \(e_5\) and \(e_1\), which place
Vera in Corvia, and the full query has a witness of \ExampleWitnessSize events.
The count is composed from the foreign-trip set, which is composed from the
country of each trip, and Vera's home country is shared by every query about
her travel. Finally, \(e_{10}\) concerns a hotel, yet it changes the count.
\end{example}

\stitle{Workload dimensions.}
We call the set of queries that a benchmark asks its \emph{workload}.
Answering a derived-state query requires three steps, which
Figure~\ref{fig:proof-dag} draws as the three layers of the running example.
The difficulty of each step can be varied
independently, so each step defines one \emph{workload dimension} of \sys.
\dstep{202}{Access} gathers the full support of the answer
(Properties~\ref{prop:implicit} and~\ref{prop:support}). \dstep{203}{Evolution} keeps the versions of
facts that are current at the cutoff. \dstep{204}{Composition} combines
intermediate states layer by layer (Property~\ref{prop:exact}). Each dimension
is measured on the gold derivation and its witnesses. The sizes of the support
and the witness, the age of the witness, and the vocabulary overlap of the
witness with the query measure Access.
The number of versions per fact and the number of obsolete events measure
Evolution. The height and width of \(\mathcal G_q\) measure Composition. \sys
samples its workload from \(C_\tau\) while controlling each dimension
(Section~\ref{sec:knobs}). Because each dimension is controlled when
queries are generated and measured on every query, a failure can be traced to
a specific dimension rather than to the query as a whole.

\subsection{Implications for Memory Systems and Evaluation}\label{sec:strategies}

\textls[-7]{A memory system can supply derived state in three ways, and the four
characteristics confront each way with a different main problem.}

\stitle{Retrieval of records.}
The system retrieves the records most similar to the query, possibly through a
graph or tree index, and leaves the derivation to a language
model~\cite{edge2024graphrag,gutierrez2024hipporag,sarthi2024raptor}. Because no record states the derived state
(Property~\ref{prop:implicit}), the records that support it share few
keywords with the query and are semantically distant from it. They are also
relevant only together, and their number can exceed any fixed retrieval
budget (Property~\ref{prop:support}). Ranking records by similarity to
the query can therefore miss the support for structural reasons, not merely
for lack of keywords.

\stitle{Materialization at ingestion.}
The system derives and stores states as events
arrive~\cite{zhong2024memorybank,xu2025amem,hu2023chatdb}. Derived state far
outnumbers recorded facts, because intermediate states combine across
entities, periods, and categories (Property~\ref{prop:exact}). The system can
therefore neither store every derived state nor know in advance which ones
will be queried. An event can also invalidate stored states that it never
mentions (Property~\ref{prop:nonmono}), so the system must find those states
and re-derive them~\cite{gupta1993dred,motik2015bf}, and stored state can lag
behind the history.

\stitle{Derivation at query time.}
The system searches and reasons when a query arrives, as agentic search
does~\cite{trivedi2023ircot}.
For every query it must gather the full support and combine every layer of
intermediate states (Properties~\ref{prop:support} and~\ref{prop:exact}), and
it recomputes intermediate states that many queries share. Each search and
derivation step involves calls to a language model, so answers come with high
latency and high token consumption.

\stitle{Implications for evaluation.}
Practical systems mix these strategies, and they place the computation in
different components with different representations. An evaluation can
therefore observe a memory system under test (SUT) only from the outside,
through the answers it gives to queries. \sys adopts this black-box view and
asks three evaluation questions.
\begin{enumerate}[label=\textbf{Q\arabic*.},leftmargin=*,itemsep=1pt]
\item \textls[-7]{\textbf{Answer reliability} (\S\ref{sec:reliability}). Does the SUT
answer correctly, with citations to its support? Missed support shows up here
first.}
\item \textbf{Failure diagnosis} (\S\ref{sec:diagnosis}). Where does a wrong
answer first depart from the gold derivation, and why? The strategies run into
problems at different steps.
\item \textbf{Derived-state maintenance} (\S\ref{sec:maintenance}). Does a
correct answer follow the history as it grows, and what does the SUT pay
at ingestion and at query time? Stored state can lag, and the strategies
place their cost at different stages.
\end{enumerate}

\section{Benchmark Construction}\label{sec:construction}\label{sec:suite}\label{sec:task}

\begin{figure*}[t]
\centering
\resizebox{0.9\textwidth}{!}{\input{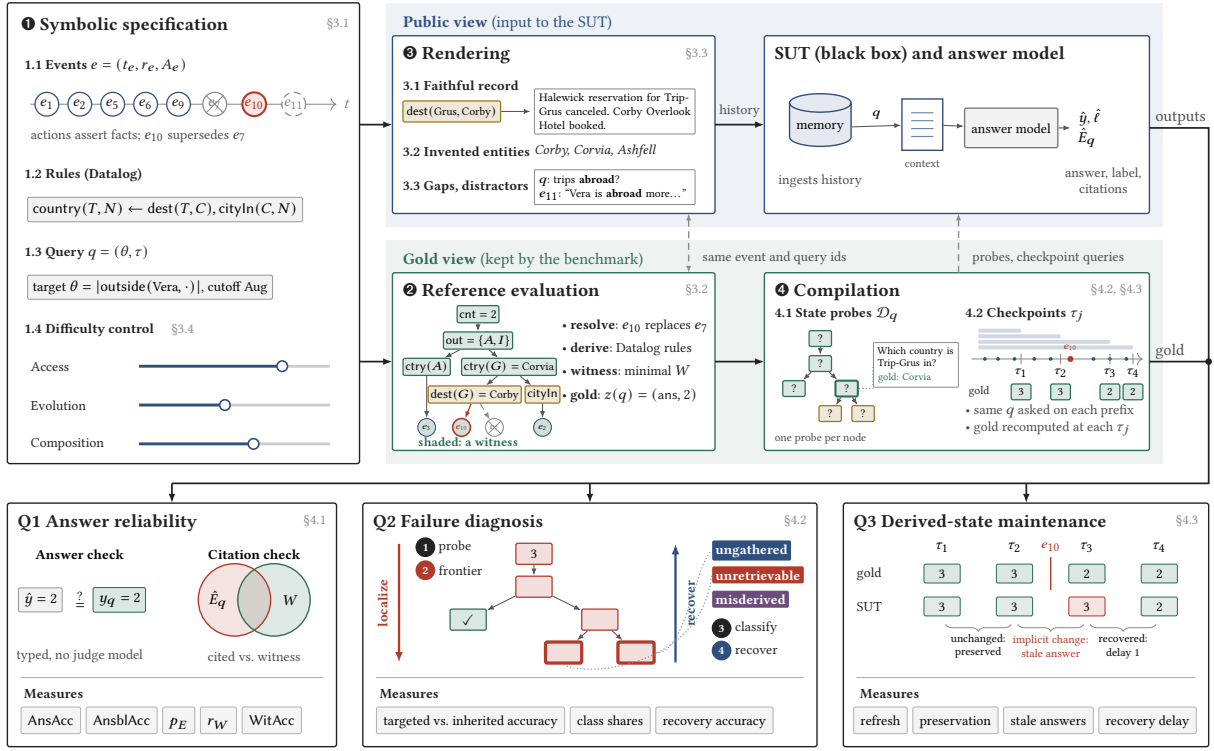}}
\Description{Construction and evaluation of one scenario. A symbolic
specification of events, rules, queries, operator families, and difficulty
controls feeds two lanes. In the lane that the SUT sees, rendering produces
faithful records with invented entities, lexical gaps, and distractors, and
the SUT ingests the history and returns a context from which the answer model
gives an answer, a label, and citations. In the lane that the benchmark keeps,
reference evaluation computes gold answers, derivation DAGs, support, and
witnesses, and compilation derives state probes, checkpoints, and gold
trajectories. The outputs feed Q1 answer reliability, Q2 failure diagnosis,
and Q3 derived-state maintenance, each with its measures.}
\caption{Construction and evaluation pipeline of \sys.}
\label{fig:pipeline}
\end{figure*}

\sys builds each scenario in three steps, shown in Figure~\ref{fig:pipeline}.
A symbolic specification (Section~\ref{sec:scenarios}) is evaluated into gold
(Section~\ref{sec:evaluator}) and then rendered into natural language
(Section~\ref{sec:rendering}). Difficulty control shapes the specification
(Section~\ref{sec:knobs}).

\subsection{Scenario Specification}\label{sec:scenarios}\label{sec:task-generation}\label{sec:operators}

Gold for derived state cannot be annotated reliably by hand, because an
answer can depend on many events and on information being absent. \sys
therefore specifies each scenario symbolically, as events with actions and
queries with targets and cutoffs. Natural language enters only at rendering,
so the specification alone determines every gold answer.

\stitle{Scenarios and domains.}
\textls[-7]{A \emph{scenario} is one evolving history together with its rules and
queries, and a \emph{domain} is a family of scenarios that share a workflow.
The suite contains \NumWorlds scenarios in \NumDomains domains. \tripops
covers travel booking and reimbursement, \shopops covers orders, returns, and
spending rules, and \eduops covers degree audits over prerequisite chains.
Each scenario adds its own rules, corrections, and cancellations to the
workflow of its domain.}

\stitle{Query targets.}
Query targets are built from operators grouped into five families, and one
target can combine several families. Each family exercises a different
construct of the derivation program, listed in Table~\ref{tab:operators}.
Appendix~\ref{sec:operator-list} lists all operators.
Because each construct demands a different computation from an SUT, success on
one family says little about another. Table~\ref{tab:operators} confirms
that the families stress different workload dimensions. Each query counts
once there, under the family of its main operation. Witness is the mean
witness size in events, Obs.\ the share of queries whose gold derivation
excludes an obsolete version, and height and width are means over the gold
derivation DAGs.

\begin{table}[t]
\caption{Operator families and the workload dimensions of their queries.}
\label{tab:operators}
\footnotesize\centering
\setlength{\tabcolsep}{3.2pt}
\begin{tabular}{@{}llrrrrr@{}}
\toprule
 & & & \multicolumn{1}{c}{Access} & \multicolumn{1}{c}{Evol.} & \multicolumn{2}{c}{Composition}\\
\cmidrule(lr){4-4}\cmidrule(lr){5-5}\cmidrule(l){6-7}
Family & Construct & Queries & Witness & Obs.\,\% & Height & Width\\
\midrule
\kw{temporal} & version selection & 191 & 1.1 & 51 & 3.0 & 2.3\\
\kw{semantic} & category mapping & 327 & 5.6 & 35 & 3.5 & 15.1\\
\kw{closure} & recursion & 208 & 2.2 & 28 & 3.3 & 7.9\\
\kw{rules} & negation & 312 & 4.1 & 65 & 3.6 & 9.8\\
\kw{set\_quant} & aggregation & 465 & 5.9 & 49 & 3.8 & 20.7\\
\midrule
All &  & 1{,}503 & 4.4 & 47 & 3.5 & 13.1\\ 
\bottomrule
\end{tabular}
\end{table}

\subsection{Reference Evaluation}\label{sec:evaluator}\label{sec:state-construction}

The \emph{reference evaluator} is an executable implementation of the
derivation program of Section~\ref{sec:closure}. One computation produces the
gold answer, the gold derivation DAG, and the witnesses of a query. While
computing the answer, the evaluator adds a node to \(\mathcal G_q\) for
everything it derives, reads, or checks, together with an edge to each input
it uses. Witnesses are extracted from the leaves of \(\mathcal G_q\) and of
alternative derivations, and each is checked to be minimal. Because all gold
objects come from this one computation, every node of \(\mathcal G_q\) has a
known correct state, which Q2 needs, and gold can be recomputed at any
cutoff, which Q3 needs.

\subsection{Natural-Language Rendering}\label{sec:rendering}\label{sec:virtualization}

Rendering produces the record \(r_e\) of each event from its actions \(A_e\)
and turns each query target into a natural-language question. The
specification determines what is correct, and rendering determines how far an
SUT must reach from the question to the records that justify the answer.

\stitle{Faithful records.}
A record carries every argument of its action, so its fact is recoverable
from the text. No record states a derived state, so
Property~\ref{prop:implicit} holds by construction.

\stitle{Entity virtualization.}
A model that recognizes an entity can answer from pretrained knowledge and
skip the records that map the entity to a category. Rendering therefore invents every recognizable entity, including people,
organizations, places, products, courses, and currencies, so that every
mapping must come from the history.

\stitle{Access in rendering.}
Questions use the vocabulary of derived predicates, and they often refer to
an entity through an associated event, a date, or a place rather than its
identifier. For some queries, the key terms of the question appear in distractor
records, which leave the answer unchanged, but in none of the supporting
records.

\subsection{Difficulty Control}\label{sec:knobs}

When generating a scenario, the generator sets how demanding each query is in
each workload dimension~\cite{erling2015ldbc}. Queries therefore spread over a range of difficulty
in every dimension, and the dimension measures of
Section~\ref{sec:dimensions} record where each query falls.

\stitle{Access.}
Support is harder to gather when its events are far apart and do not resemble
the question. Through \emph{dispersion}, the generator spreads the events of
one witness over the history, and rendering adds the lexical gaps, indirect
references, and distractors of Section~\ref{sec:rendering}.

\stitle{Evolution.}
Current versions are harder to identify when facts are revised often and
obsolete records appear late. The generator builds \emph{update chains} that
revise a fact several times and places obsolete records late in the history.

\stitle{Composition.}
A query is harder to compose when it combines more kinds of reasoning. The
generator varies this number, the \emph{composition degree}, from a single
operation to combinations of version resolution, rule application, and
checking that a collection is complete.

\stitle{Quota bands.}
Target shares of operator families, witness sizes, and answer types keep the
workload off the easiest cases.

\subsection{Suite Statistics}\label{sec:suite-stats}

Table~\ref{tab:suite} summarizes the ten scenarios. Because scenarios are
generated from configurations, new scenarios of the same design can be
generated whenever existing ones may have entered the training data of a
model.

\begin{table}[t]
\caption{Suite statistics by domain. Scen.\ is the number of scenarios.
Labels are \lblans, \lblzero, and \lblunans. Witness sizes are medians and
maxima in events.}
\label{tab:suite}
\footnotesize\centering
\setlength{\tabcolsep}{3pt}
\begin{tabular}{@{}lrrrrrrrr@{}}
\toprule
 & & & & \multicolumn{3}{c}{Labels} & \multicolumn{2}{c}{Witness}\\
\cmidrule(lr){5-7}\cmidrule(l){8-9}
Domain & Scen. & Events & Queries & Ans. & Zero & Unk. & Med. & Max\\
\midrule
\tripops & 4 & 1{,}817 & 561 & 409 & 77 & 75 & 3 & 69\\
\shopops & 3 & 1{,}462 & 473 & 344 & 42 & 87 & 2 & 31\\
\eduops & 3 & 1{,}413 & 469 & 423 & 18 & 28 & 3 & 20\\
\midrule
All & 10 & 4{,}692 & 1{,}503 & 1{,}176 & 137 & 190 & 3 & 69\\ 
\bottomrule
\end{tabular}
\end{table}

\section{Evaluation Methodology}\label{sec:methodology}

The protocol follows the black-box view of Section~\ref{sec:strategies}. For
every query, an SUT returns a \emph{context}, the text it supplies for
answering, and one answer model shared by all SUTs turns it into an answer, an
answerability label, and cited events. Every measurement below is obtained by
posing questions to the SUT.

\subsection{Q1: Answer Reliability}\label{sec:reliability}\label{sec:scoring}

Q1 asks whether an SUT answers correctly and cites its support. A correct
answer alone is weak evidence, because Booleans and empty sets can be guessed
and a total can be right by coincidence.

\stitle{Answer scoring.}
Answers are scored deterministically. Every gold answer has a type, and each
type has its own equivalence test for numbers, Booleans, collections, or
accepted aliases. No language model is therefore needed to judge an answer,
and scores are fully reproducible. Answer accuracy (\ansacc) requires the
returned answer to match \(y_q\) for an \lblans or \lblzero query and the
\lblunans label otherwise. Answerability accuracy (\lblacc) scores the label alone.

\stitle{Citation scoring.}
\sys also scores the cited events \(\hat E_q\) against the support and
witnesses of Section~\ref{sec:provenance}:
\begin{equation}\label{eq:evidence-pr}
p_E(q)=\frac{|\hat E_q\cap S(q)|}{|\hat E_q|},
\qquad
r_W(q)=\max_{W\in\mathcal W(q)}\frac{|\hat E_q\cap W|}{|W|}.
\end{equation}
\textls[-2]{Their harmonic mean is the citation \(F_1\). Witnessed accuracy (\witacc) counts a query as correct when its answer is
correct and both \(p_E\) and \(r_W\) reach \(0.5\).} A correct answer then counts
only if most of its citations are relevant and cover a substantial part of one
witness, without requiring a complete citation list. The two scores encode two
design decisions. Precision checks citations against the whole support \(S(q)\), so citing additional supporting events costs nothing. Recall takes
the best match over all witnesses, so an SUT that follows a
different valid derivation loses nothing. The shares of queries that cite an obsolete event or a distractor complement
both scores.

\subsection{Q2: Failure Diagnosis}\label{sec:diagnosis}

Q2 asks where a wrong answer arises and why. Because SUTs are black boxes,
diagnosis cannot inspect their internal state. It can only ask further
questions, and \(\mathcal G_q\) tells it which questions to ask and what their
correct answers are.

\subsubsection{Failure Classes}\label{sec:failure-classes}
A wrong answer arises at some state of the derivation. The SUT may be able
to produce that state but fail to gather it when the derivation needs it, or
it may fail to produce the state even when asked for it directly. Each case
calls for a different repair, so \sys distinguishes three \emph{failure
classes}.
\begin{itemize}[leftmargin=1em,itemsep=1pt]
\item A state is \emph{ungathered} when the SUT returns it when asked for it
directly, but the request for the state that needs it does not gather it.
\item A resolved fact or rule is \emph{unretrievable} when the SUT cannot
return it even when asked for it directly.
\item A derived state is \emph{misderived} when the SUT derives it wrongly even
when asked for it directly, although its inputs are gathered.
\end{itemize}
The classes follow the workload dimensions of Section~\ref{sec:dimensions}.
Ungathered states are failures of Access. Unretrievable facts arise at Access
or Evolution, depending on whether the fact has a superseded version, and
misderived states arise at Composition.

\subsubsection{State Probes}\label{sec:probes}
Any node of \(\mathcal G_q\) that holds a resolved fact, a rule, or a derived
state can be asked about directly.

\begin{definition}[State probe]\label{def:state-probe}
A \emph{state probe} of a node \(v\) in \(\mathcal G_q\) is a natural-language
question at the cutoff of \(q\) whose gold answer is the state of \(v\).
\end{definition}

Probes are rendered from fixed templates, and their gold answers come from the
reference evaluator, so no language model writes or grades them. The suite
contains \NumStateProbes probes for \NumProbeRoots queries. The probed nodes,
connected by the dependencies of \(\mathcal G_q\), form the \emph{diagnostic
DAG} \(\mathcal D_q\). Every SUT answers the same probes for the same
queries in the same order.

\stitle{Two contexts.}
\textls[-6]{Each probe is answered under two contexts. In the \emph{inherited} context,
the answer model receives the context that the SUT returned for the question
of a parent node, which for the children of the root \(r_q\) is the original
query.
This reproduces what the derivation had available when it needed the state.
For the root, the inherited answer is the original answer. In the
\emph{targeted} context, the probe itself is sent to the SUT as a new
query, and the answer model receives the context returned for it. This tests
whether the SUT can return the information when asked for it directly. Let
\(t_s(v)\in\{0,1\}\) indicate whether SUT \(s\) answers the probe of \(v\)
correctly in the targeted context. Because a node can have several parents,
inherited correctness belongs to an edge. For a child \(u\) of \(v\),
\(i_s(u\mid v)\in\{0,1\}\) indicates whether \(s\) answers the probe of \(u\)
correctly in the context returned for the question of \(v\), and \(i_s(r_q)\)
indicates whether the original answer is correct.}

\subsubsection{Failure Localization}\label{sec:localization}
A wrong probe answer does not by itself locate a failure, because an error can
propagate upward, and a node computed correctly from a wrong input is wrong as
well. The failure arises at a node that is wrong although all of its inputs
are right. Such nodes mark where an SUT first departs from the gold
derivation, and together they form its \emph{failure frontier}. Localization
uses the inherited context, because it shows whether each request gathered
the states it needed.

\begin{definition}[Failure frontier]\label{def:frontier}
A probed node \(v\) lies on the \emph{failure frontier} of SUT \(s\) when
some path \(r_q=v_0,\dots,v_k=v\) in \(\mathcal D_q\) has \(i_s(r_q)=0\) and
\(i_s(v_j\mid v_{j-1})=0\) for every \(j\), and \(i_s(u\mid v)=1\) for every
child \(u\) of \(v\).
\end{definition}

A derivation can have one frontier node per failing branch. The targeted
context determines the class of a frontier node \(v\). If \(t_s(v)=1\), the SUT
returns \(v\) when asked directly, so \(v\) is ungathered. Otherwise \(v\) is
unretrievable if it is a resolved fact or rule, and misderived if it is a
derived state, because the inputs of \(v\) were gathered.
Algorithm~\ref{alg:proof-guided-diagnosis} locates the frontier and classifies
each of its nodes. It uses only probe answers, so diagnosis needs no access to
the internals of the SUT.

\begin{algorithm}[t]
\small
\caption{Failure frontier localization}
\label{alg:proof-guided-diagnosis}
\KwIn{diagnostic DAG \(\mathcal D_q\), correctness \(i_s\) and \(t_s\) of SUT \(s\)}
\KwOut{frontier nodes with their failure classes}
\lIf{\(i_s(r_q)=1\)}{\Return{\(\varnothing\)}}
\(F\leftarrow\varnothing\);\ \(V\leftarrow\varnothing\);\ \(K\leftarrow[\,r_q\,]\)\;
\While{\(K\neq\varnothing\)}{
 \(v\leftarrow\) pop \(K\); \textbf{if} \(v\in V\) \textbf{then continue}; \(V\leftarrow V\cup\{v\}\)\;
 \(U\leftarrow\{u\in\kw{children}(v) : i_s(u\mid v)=0\}\)\;
 \uIf(\tcp*[f]{error from below}){\(U\neq\varnothing\)}{push \(U\) onto \(K\)}
 \uElseIf{\(t_s(v)=1\)}{\(F\leftarrow F\cup\{(v,\text{ungathered})\}\)}
 \uElseIf{\(v\) is a derived state}{\(F\leftarrow F\cup\{(v,\text{misderived})\}\)}
 \lElse{\(F\leftarrow F\cup\{(v,\text{unretrievable})\}\)}
}
\Return{\(F\)}\;
\end{algorithm}

\subsubsection{Interventions}\label{sec:interventions}
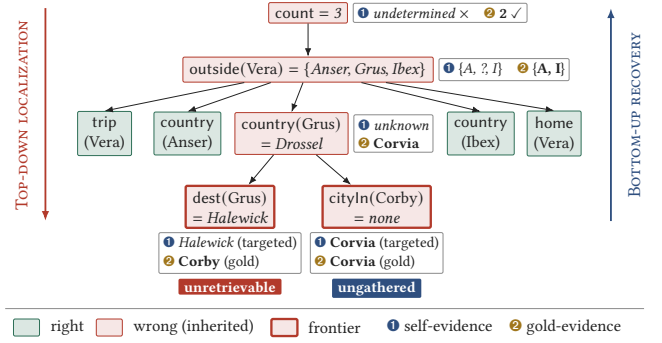
\begin{figure}[t]
\centering
\resizebox{\columnwidth}{!}{%
\begin{tikzpicture}[font=\small,
 /utils/exec={%
   \definecolor{pink}{HTML}{222222}%
   \definecolor{pgray}{HTML}{8A8A8A}%
   \definecolor{pcomp}{HTML}{2E6F57}%
   \definecolor{pevol}{HTML}{A0761C}%
   \definecolor{pacc}{HTML}{2F4F7F}%
   \definecolor{paccent}{HTML}{B23A2B}},
 node/.style={line width=0.55pt,rounded corners=1pt,align=center,
   inner xsep=4pt,inner ysep=3pt,font=\large,text=pink},
 ok/.style={node,draw=pcomp,fill=pcomp!17},
 bad/.style={node,draw=paccent,fill=paccent!12},
 front/.style={bad,line width=1.4pt},
 rec/.style={draw=pgray,line width=0.4pt,fill=white,rounded corners=1pt,
   align=left,inner sep=2.5pt,font=\normalsize,text=pink},
 tag/.style={font=\normalsize\bfseries,text=white,inner xsep=3.5pt,inner ysep=1.5pt,
   line width=0pt,rounded corners=1pt},
 arr/.style={-{Latex[length=5pt,width=3.8pt]},draw=pink,line width=0.55pt},
 dir/.style={-{Latex[length=6pt,width=4.6pt]},line width=1.1pt,draw=#1},
 dirlabel/.style={rotate=90,font=\Large\scshape,text=#1},
 key/.style={font=\large,anchor=west,text=pink}]
\newcommand{\one}{\textcolor{pacc}{\ding{202}}}
\newcommand{\two}{\textcolor{pevol}{\ding{203}}}
\node[bad] (root) at (0,0) {\(\kw{count}=\textit{3}\)};
\node[bad] (out) at (0,-1.2) {\(\kw{outside}(\text{Vera})=\{\textit{Anser},\textit{Grus},\textit{Ibex}\}\)};
\node[ok] (trips) at (-4.3,-2.55) {\(\kw{trip}\)\\(Vera)};
\node[ok] (ca) at (-2.55,-2.55) {\(\kw{country}\)\\(Anser)};
\node[bad] (cg) at (-0.35,-2.55) {\(\kw{country}(\text{Grus})\)\\\(=\textit{Drossel}\)};
\node[ok] (ci) at (3.65,-2.55) {\(\kw{country}\)\\(Ibex)};
\node[ok] (home) at (5.2,-2.55) {\(\kw{home}\)\\(Vera)};
\node[front] (dg) at (-1.65,-4.1) {\(\kw{dest}(\text{Grus})\)\\\(=\textit{Halewick}\)};
\node[front] (cc) at (1.45,-4.1) {\(\kw{cityIn}(\text{Corby})\)\\\(=\textit{none}\)};
\draw[arr] (root) -- (out);
\foreach \n in {trips,ca,cg,ci,home} \draw[arr] (out) -- (\n.north);
\draw[arr] (cg) -- (dg);
\draw[arr] (cg) -- (cc);
\node[rec,anchor=west] at ([xshift=3pt]root.east) {\one\ \textit{undetermined} \(\times\)\quad\two\ \textbf{2} \(\checkmark\)};
\node[rec,anchor=west] at ([xshift=3pt]out.east) {\one\ \{\textit{A, ?, I}\}\quad\two\ \{\textbf{A, I}\}};
\node[rec,anchor=west] at ([xshift=3pt]cg.east) {\one\ \textit{unknown}\\\two\ \textbf{Corvia}};
\node[rec,anchor=north] (rdg) at ([yshift=-2pt]dg.south) {\one\ \textit{Halewick} (targeted)\\\two\ \textbf{Corby} (gold)};
\node[rec,anchor=north] (rcc) at ([yshift=-2pt]cc.south) {\one\ \textbf{Corvia} (targeted)\\\two\ \textbf{Corvia} (gold)};
\node[tag,fill=paccent,anchor=north] at ([yshift=-3pt]rdg.south) {unretrievable};
\node[tag,fill=pacc,anchor=north] at ([yshift=-3pt]rcc.south) {ungathered};
\draw[dir=paccent] (-5.55,0.1) -- (-5.55,-4.4);
\node[dirlabel=paccent] at (-6.05,-2.15) {Top-down localization};
\draw[dir=pacc] (6.4,-4.4) -- (6.4,0.1);
\node[dirlabel=pacc] at (6.9,-2.15) {Bottom-up recovery};
\draw[pgray,line width=0.4pt] (-6.4,-6.25) -- (7.1,-6.25);
\node[ok,minimum width=0.55cm,minimum height=0.36cm] at (-5.95,-6.65) {};
\node[key] at (-5.55,-6.65) {right};
\node[bad,minimum width=0.55cm,minimum height=0.36cm] at (-4.2,-6.65) {};
\node[key] at (-3.8,-6.65) {wrong (inherited)};
\node[front,minimum width=0.55cm,minimum height=0.36cm] at (-0.5,-6.65) {};
\node[key] at (-0.1,-6.65) {frontier};
\node[key] at (1.55,-6.65) {\one\ self-evidence\quad\two\ gold-evidence};
\end{tikzpicture}}
\Description{Diagnosis of the running example on one DAG. Node colors show
the probe answers in the inherited context: the count, the foreign-trip set,
and the country of Trip-Grus are wrong. Localization runs top-down to two
frontier nodes: the destination of Trip-Grus, answered Halewick also in the
targeted context and labeled unretrievable, and the country of Corby, missing
in the inherited context but answered Corvia in the targeted context and
labeled ungathered. Tags beside each failing node show the states that
self-evidence recovery and gold-evidence recovery compute bottom-up: self
evidence leaves the country of Trip-Grus unknown and the count wrong, gold
evidence yields Corvia and a count of 2.}
\vspace{-4pt}
\caption{Diagnosis of Example~\ref{ex:budget-diagnosis}.}
\label{fig:q2-diagnosis}
\end{figure}

A frontier shows that a state is wrong while the answer is wrong, which is a
co-occurrence rather than an explanation. Two \emph{recoveries} test whether
the departure matters for the answer. Both rebuild the answer bottom-up along
\(\mathcal D_q\). Each branch is followed down to a node that the SUT answers
correctly in the targeted context, or to a leaf, and the answer model
recomputes every node above these stopping nodes from its children. The two
recoveries differ only in the states at the stopping nodes. Figure~\ref{fig:q2-diagnosis} marks their results with \dnum{202} and
\dnum{203}.

\stitle{\dstep{202}{Self-evidence recovery.}}
The stopping nodes keep the SUT's targeted answers, so every state comes from
the SUT. Success shows that the SUT can return the evidence the answer needs
and that the original failure lay in gathering and combining it.

\stitle{\dstep{203}{Gold-evidence recovery.}}
Wrong stopping nodes, which are leaves, take their gold states instead, and
everything else stays as the SUT produced it. Success where self-evidence
recovery fails shows that the wrong evidence causes the failure. We call such
evidence \emph{load-bearing}.

\stitle{Discussion.}
The two recoveries complement the failure classes. Self-evidence recovery can
repair an ungathered failure, because the SUT returns the state when asked
directly. Gold-evidence recovery can also repair an unretrievable one. A
failure that persists after gold-evidence recovery lies above the evidence,
in how states are combined. Both recoveries measure how much of a failure can be recovered with
assistance.

\begin{example}[Diagnosis of Example~\ref{ex:travel-state}]\label{ex:budget-diagnosis}
An SUT answers three instead of two. In Figure~\ref{fig:q2-diagnosis}, node
colors show its probe answers in the inherited context. Vera's home country, the set of trips,
and the countries of Trip-Anser and Trip-Ibex are right, but Trip-Grus is placed
in Drossel. Both inputs of that state are wrong, so both lie on the frontier.
The destination of Trip-Grus is Halewick from the obsolete \(e_7\), also in the
targeted context, so this resolved fact is unretrievable, and its dimension is
Evolution. The country of Corby is missing from the context of its parent, but
the targeted context returns Corvia, so this fact is ungathered.
The tags beside the failing nodes show what each recovery computes from the
stopping nodes upward. Self-evidence recovery
repairs the ungathered fact, but Halewick contradicts it, so the country
of Trip-Grus stays unknown and the count is still wrong.
Gold-evidence recovery replaces Halewick with Corby and yields two, so the
obsolete destination is load-bearing.
\end{example}

\subsection{Q3: Derived-State Maintenance}\label{sec:maintenance}

\textls[-1]{Q3 asks whether an SUT keeps derived state correct as the history grows. By Property~\ref{prop:nonmono}, an answer that is correct at one cutoff need
not remain correct at the next, and an SUT that materializes state may lag
behind the history. One snapshot
cannot measure maintenance.}

\stitle{Fixed-query trajectories.}
The evaluation fixes a sequence of cutoffs \(\tau_1<\cdots<\tau_m\), called
\emph{checkpoints}, spread over the history. A \emph{tracked query} fixes its
target \(\theta\) and is asked at every checkpoint, and
\(q_j=(\theta,\tau_j)\) denotes its instance at checkpoint \(j\). The
reference evaluator recomputes gold at each checkpoint, which yields the
\emph{trajectory} \(z(q_1),\ldots,z(q_m)\). Before each checkpoint, the SUT
ingests the new events into its existing memory, so it must update that memory
rather than rebuild it.

\stitle{Refresh and preservation.}
Consecutive checkpoints \(j-1\) and \(j\) form a \emph{transition}. Let
\(d_j(q)=\mathbf 1[z(q_j)\neq z(q_{j-1})]\), and call the transition
\emph{changed} when \(d_j(q)=1\) and unchanged otherwise. Tracked queries are
sampled so that both kinds of transition occur often. Refresh measures whether
a change reaches the answer, and preservation measures whether an unaffected
answer survives. With \(c_{s,j}(q)\) indicating that SUT \(s\) answers
\(q_j\) correctly,
\begin{equation}\label{eq:maintenance-rates}
\begin{aligned}
\operatorname{Refresh}_s
 &=\frac{\sum_{q,j}d_j(q)\,c_{s,j-1}(q)\,c_{s,j}(q)}
         {\sum_{q,j}d_j(q)\,c_{s,j-1}(q)},\\
\operatorname{Preservation}_s
 &=\frac{\sum_{q,j}(1-d_j(q))\,c_{s,j-1}(q)\,c_{s,j}(q)}
         {\sum_{q,j}(1-d_j(q))\,c_{s,j-1}(q)}.
\end{aligned}
\end{equation}
Both rates condition on a correct previous answer. The condition separates
maintenance from initial difficulty, since an SUT that never answered
correctly has nothing to maintain.

\begin{definition}[Explicit and implicit change]\label{def:implicit-change}
A changed transition at checkpoint \(j\) is \emph{explicit} if some event \(e\)
with \(\tau_{j-1}<t_e\le\tau_j\) belongs to a witness of \(q_j\) and an action
in \(A_e\) asserts a fact that contains the new answer. Otherwise the change
is \emph{implicit}.
\end{definition}

An implicit change is the case that Property~\ref{prop:nonmono} describes,
because no arriving record mentions the state that changed. Comparing refresh
across the two kinds shows whether an SUT propagates a change through
dependencies or only updates what it is told. The latter is the lag that
Section~\ref{sec:strategies} attributes to materialization. In Example~\ref{ex:travel-state},
\(e_{10}\) makes an implicit change: it books a hotel, yet the count drops
from three to two.

\stitle{Complementary measures.}
Four further measures complement the two rates.
\begin{itemize}[leftmargin=1em,itemsep=1pt]
\item \emph{Stale-answer rate.} The share of changed transitions at which the
SUT still returns the previous gold answer. It distinguishes a missed
update from a wrong one.
\item \emph{Recovery delay.} The number of checkpoints from a changed
transition until the SUT first answers correctly.
\item \emph{Continuous correctness.} The share of tracked queries answered
correctly at every checkpoint up to \(\tau_j\). It combines initial difficulty
with maintenance.
\item \emph{Ingestion and query-time cost.} The cost of ingesting events and
of answering queries, recorded separately. It shows where each SUT places
its work among the strategies of Section~\ref{sec:strategies}.
\end{itemize}

\section{Experiments}\label{sec:results}
\setlength{\textfloatsep}{10pt plus 4pt minus 3pt}
\setlength{\floatsep}{8pt plus 4pt minus 2pt}

\subsection{Experimental Setup}\label{sec:experimental-setup}

\stitle{SUTs and reference conditions.}
The SUTs are four memory systems and two raw-retrieval baselines. The memory
systems are \memzero~\cite{chhikara2025mem0}, \cognee~\cite{markovic2025cognee},
\lightmem{}~\cite{fang2026lightmem}, and \memoryos{}~\cite{kang2025memoryos}.
They cover extracted-fact stores, graph-oriented memory, and hierarchical
organizations. The baselines rank the history with \bmtf{} or with dense
embeddings (\embed{}) and return the top items. They retrieve session
documents by default, and figures mark session-level and event-level variants
as -S and -E where both appear. Three reference conditions bound the results.
\closedbook receives no history, \fullmode receives the whole visible history,
and \oraclemode receives the support \(S(q)\), all events that contribute to
the answer.

\stitle{Models and defaults.}
\textls[-7]{The answer model is \answerer{} with reasoning disabled and temperature
zero. All SUTs share our answer and judge prompts, released with the artifact.
The judge prompt is stricter than those used in the original benchmarks, so
scores can differ from previously reported ones. Memory systems that extract memories at
ingestion use \textsf{GPT-5.6 Luna} as the extraction model. Dense retrieval
uses \textsf{Qwen3-Embedding-0.6B}~\cite{zhang2025qwen3}. Every SUT returns \(k=15\) items per
query. \memzero and \lightmem{} do not return event identifiers, so their
citations are mapped to events through the recorded memory attribution.}

\stitle{Query sets.}
The main comparison asks all \NumQueriesTotal{} queries of the \NumWorlds{}
scenarios at the end of each history, over \NumEventsTotal{} events. The
context-budget comparison caps the context at 500 to 16K tokens on
\NumContextQueries{} queries. Diagnosis probes the complete diagnostic DAGs of
a subset of queries from every scenario, the diagnostic queries. Maintenance tracks
\NumTrajectoryQueries{} queries at five checkpoints, at 20\%, 40\%, 60\%, 80\%,
and 100\% of each history. Transition analysis uses the tracked queries whose
gold is never \lblunans, which contain \NumChangedPairs{} changed transitions.
For comparison, the SUTs are also evaluated on LoCoMo and
LongMemEval~\cite{maharana2024locomo,wu2025longmemeval} with their official
question sets. The artifact also records model settings, per-query outputs,
and the inputs of every figure.

\newcommand{\capabove}{3pt}
\newcommand{\capbelow}{3pt}
\subsection{Q1: Answer Reliability}\label{sec:results-reliability}\label{sec:results-workload}

\begin{figure}[t]
\centering
\includegraphics[width=\columnwidth]{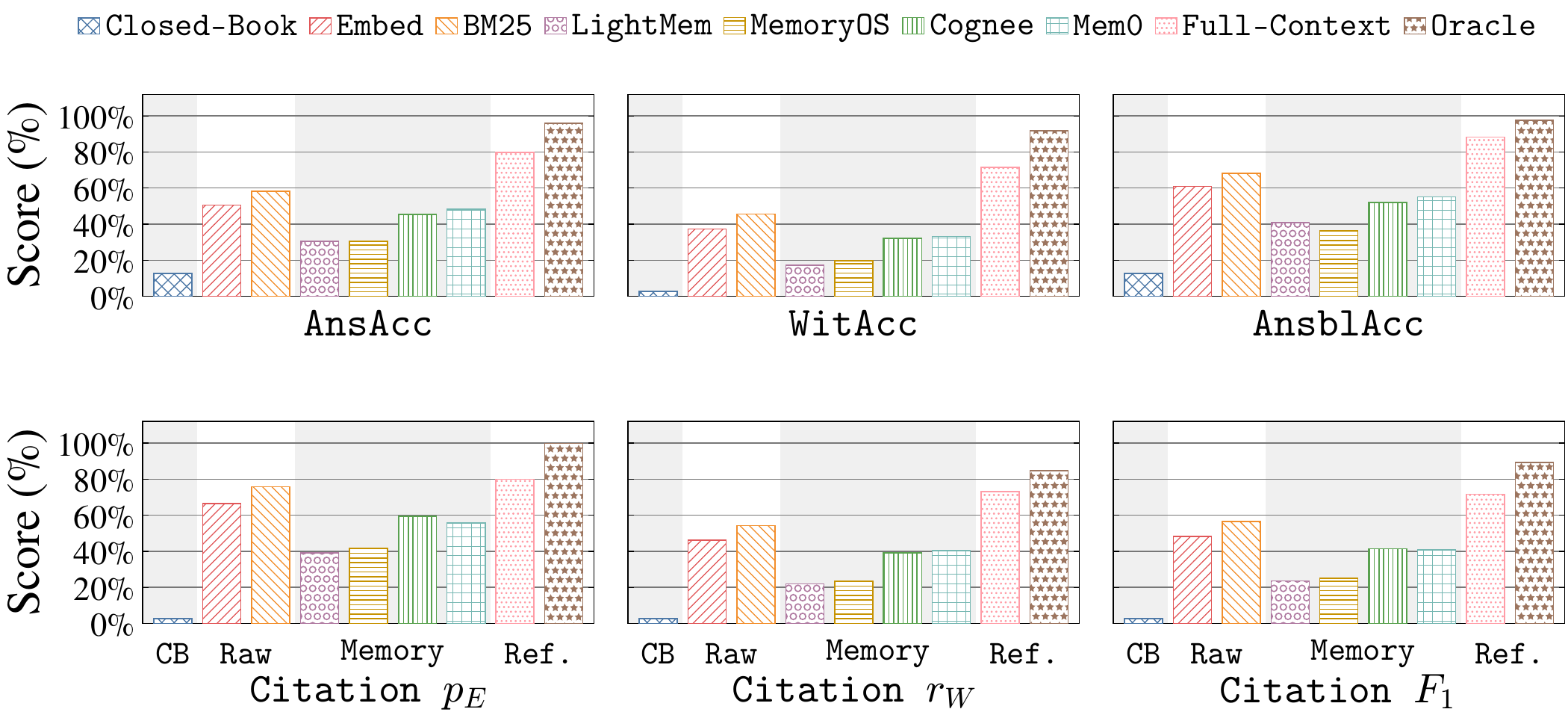}
\Description{Six panels compare answer accuracy, witnessed accuracy,
answerability accuracy, citation precision, citation recall, and citation F1
for the SUTs and reference conditions.}
\caption{Answer and citation quality on all queries.}
\label{fig:e1-ladder-metrics}
\end{figure}

\stitle{Memory systems trail raw retrieval.}
Figure~\ref{fig:e1-ladder-metrics} reports the main comparison. The four
memory systems reach \EOneMemoryAnsBand{}\% answer accuracy, and even the
strongest, \memzero, trails \bmtf{} at \EOneBMAns{}\%. \fullmode and
\oraclemode reach \EOneFullAns{}\% and \EOneOracleAns{}\% under the same
answer model, so the gap lies in what the SUTs supply. Witnessed accuracy,
answerability accuracy, and citation precision, recall, and \(F_1\) show the
same order.

\begin{figure}[t]
\centering
\includegraphics[width=\columnwidth]{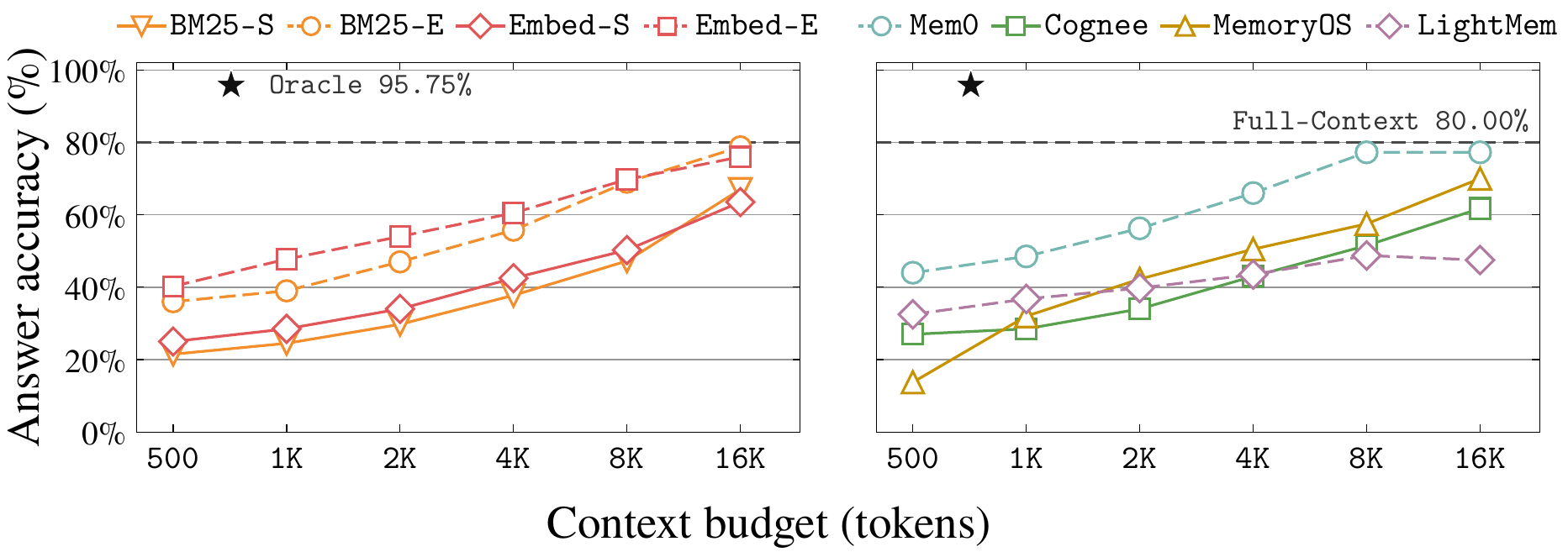}
\Description{Answer accuracy as the context budget grows, for session- and
event-level retrieval (left) and four memory systems (right).}
\caption{Answer accuracy under context budgets.}
\label{fig:q1-context-budget}
\end{figure}

\stitle{The gap persists under matched context budgets.}
Figure~\ref{fig:q1-context-budget} caps the context of every SUT at the same
number of tokens. \memzero plateaus at \EFourMemEight{}\% from 8K tokens, and
at 16K event-level \bmtf{} still leads with \EFourBMHigh{}\%. A larger budget
therefore does not favor memory over raw retrieval. With the whole history,
medium reasoning raises accuracy from \EFourFull{}\% to \EFourFullMedium{}\%,
so reasoning also matters once all evidence is present. \oraclemode needs
only 712 tokens on average.

\begin{figure}[t]
\centering
\includegraphics[width=\columnwidth]{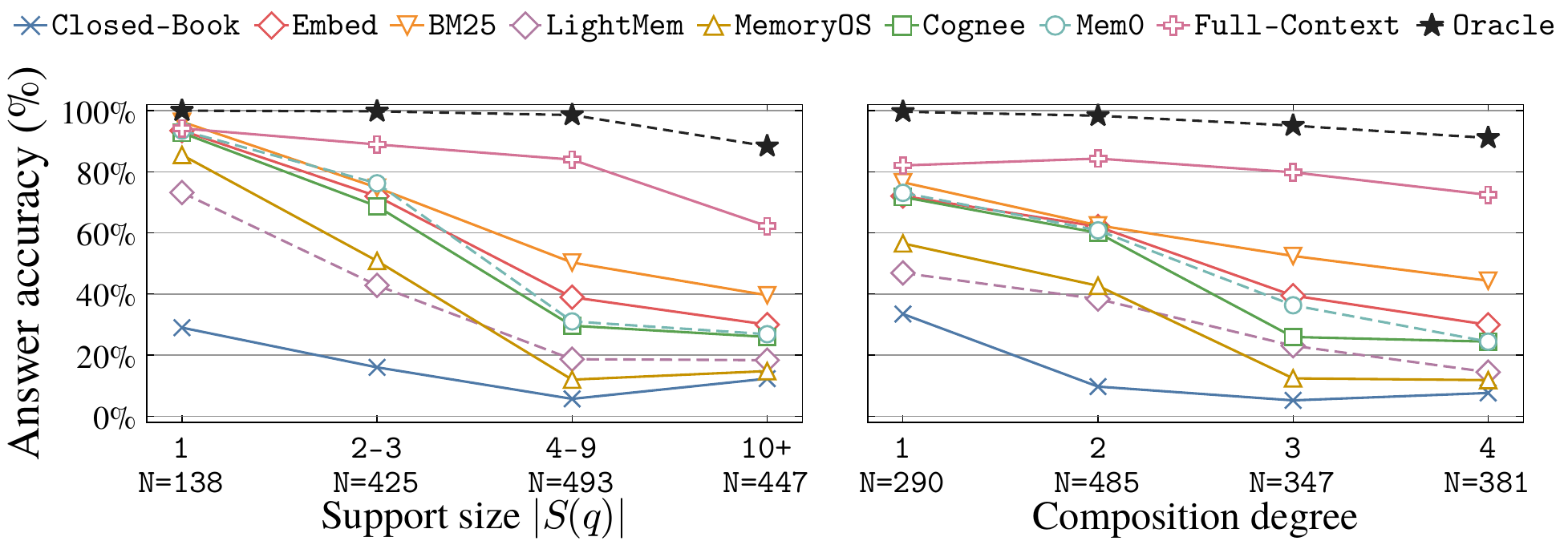}\par
{\small\textbf{(a) Access and Composition}}\par\smallskip
\includegraphics[width=\columnwidth]{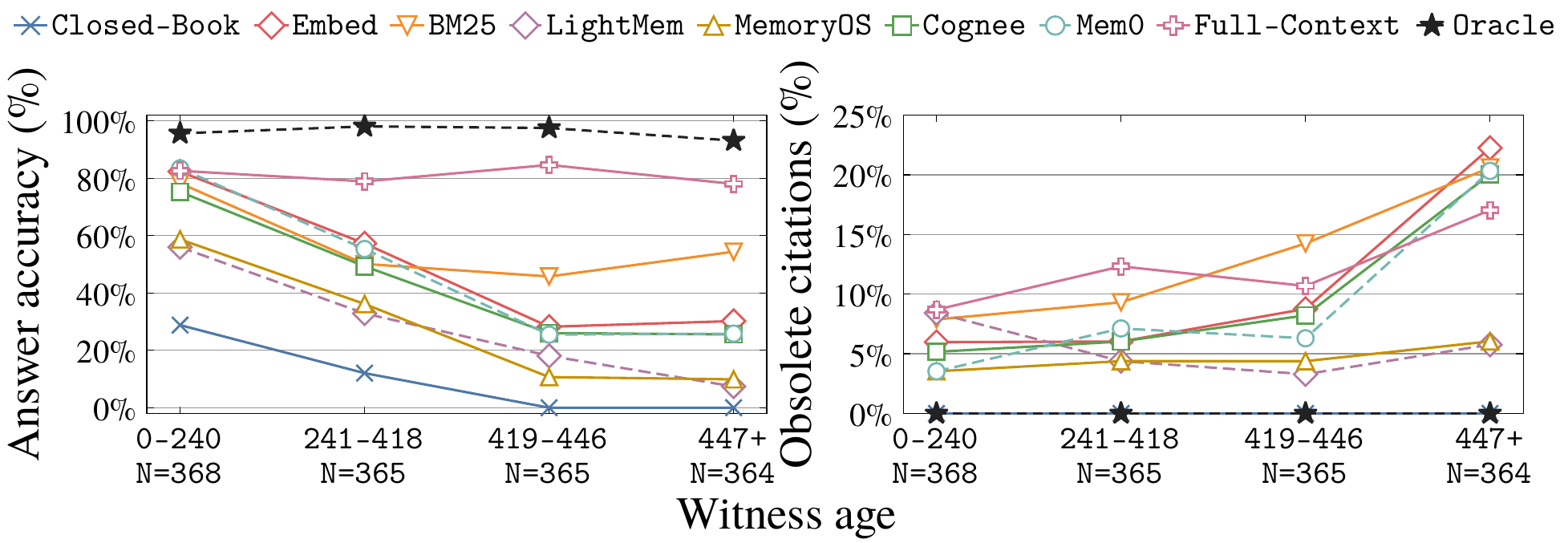}\par
{\small\textbf{(b) Access and Evolution}}
\Description{Answer accuracy by support size and composition degree,
followed by answer accuracy and obsolete-citation rate over quartiles of
witness age.}
\caption{Answer accuracy and obsolete citations along the workload
dimensions.}
\label{fig:q2-failure-strata}
\end{figure}

\begin{figure}[t]
\centering
\includegraphics[width=\columnwidth]{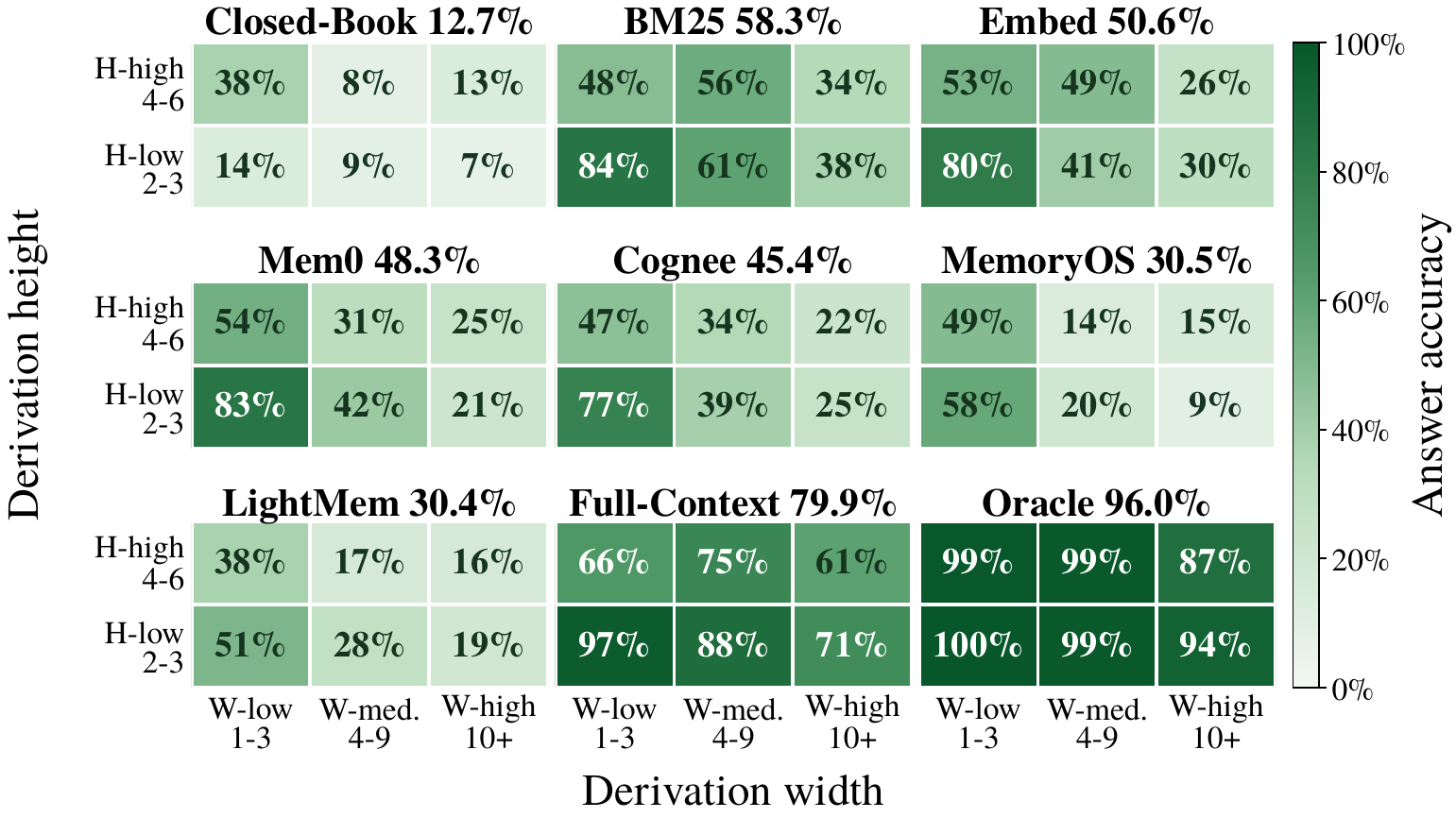}
\Description{Answer accuracy for nine conditions grouped by the height and
width of the gold derivation DAG.}
\caption{Answer accuracy by the height and width of \(\mathcal G_q\).}
\label{fig:q2-dag-height-width}
\end{figure}

\begin{figure}[t]
\centering
\includegraphics[width=\columnwidth]{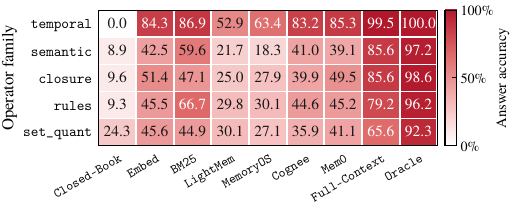}
\Description{Answer accuracy grouped by the operator family of each query.}
\caption{Answer accuracy by operator family.}
\label{fig:q2-query-families}
\end{figure}

\stitle{Accuracy falls along every workload dimension.}
Figures~\ref{fig:q2-failure-strata} and~\ref{fig:q2-dag-height-width} split
the main comparison by the measures of Section~\ref{sec:dimensions}.
For Access, memory systems fall from \EThirteenSourcesOneBand{}\% when the
support holds one event to \EThirteenSourcesManyBand{}\% when it holds ten or
more, and \memzero falls from \EThirteenMemNear{}\% to \EThirteenMemFar{}\%
from the nearest to the farthest quartile of witness age. For Evolution,
\EThirteenMemFarStale{}\% of \memzero's answers in that farthest quartile cite
an obsolete event. For Composition, accuracy falls with the composition degree
and, even among shallow derivations, with width, for \memzero from
\ShapeMemZeroLow{}\% to \ShapeMemZeroHigh{}\%. On high and wide derivations,
\oraclemode still reaches \ShapeOracleHighWide{}\%, whereas memory systems
reach \ShapeMemoryHighWideBand{}\%, so width matters alongside height. Across operator
families (Figure~\ref{fig:q2-query-families}), set and quantitative queries
are the hardest.

\begin{figure}[t]
\centering
\includegraphics[width=\columnwidth]{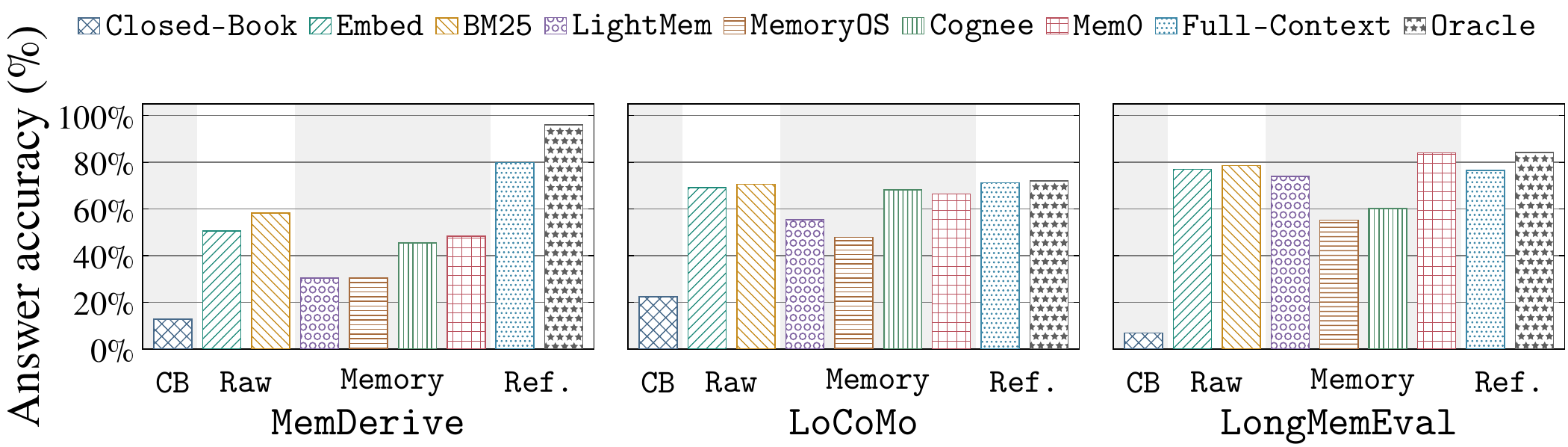}
\Description{Answer accuracy on MemDerive, LoCoMo, and LongMemEval,
shown in separate panels for the same conditions.}
\caption{Answer accuracy on three memory benchmarks.}
\label{fig:e8-cross-benchmark}
\end{figure}

\stitle{Rankings on existing benchmarks do not carry over.}
Figure~\ref{fig:e8-cross-benchmark} compares the same conditions on LoCoMo,
LongMemEval, and \sys. On LongMemEval, \memzero reaches \EEightLMEMem{}\% and
exceeds \bmtf{} at \EEightLMEBM{}\%. On LoCoMo, \cognee reaches
\EEightLoCoMoCognee{}\%, close to \bmtf{} at \EEightLoCoMoBM{}\%. On \sys, \bmtf{}
exceeds all four memory systems. Each benchmark keeps its own question set,
so the comparison holds within each panel.

\resultfinding{1}{Memory systems answer \EOneMemoryAnsBand{}\% of
derived-state queries and trail \bmtf{}, also under matched context budgets.
Accuracy falls along all three workload dimensions, and rankings on existing
memory benchmarks do not predict these results.}

\subsection{Q2: Failure Diagnosis}\label{sec:results-diagnosis}

Diagnosis runs Algorithm~\ref{alg:proof-guided-diagnosis} on the complete
diagnostic DAGs for the raw-retrieval baselines and the four memory systems.
Table~\ref{tab:q2-classes} reports the \emph{targeted accuracy}, the share of
state probes that an SUT answers correctly in the targeted context, for four
kinds of node. They are resolved facts, rules, completeness facts, which assert
that a collection is complete, and derived states. The table also reports the
failure classes and both recoveries as shares of wrong answers. A wrong answer
can contain frontier nodes of several classes.

\begin{table}[t]
\caption{Failure diagnosis of the six SUTs (\%).}
\label{tab:q2-classes}
\footnotesize\centering
\setlength{\tabcolsep}{2.6pt}
\renewcommand{\arraystretch}{0.9}
\begin{tabular}{@{}lrrrrrrrrr@{}}
\toprule
 & \multicolumn{4}{c}{Targeted accuracy} & \multicolumn{3}{c}{Failure class} & \multicolumn{2}{c}{Recovery}\\
\cmidrule(lr){2-5}\cmidrule(lr){6-8}\cmidrule(l){9-10}
SUT & Fact & Rule & Compl. & Deriv. & Ung. & Unr. & Mis. & Self & Gold\\
\midrule
\bmtf{} & 94 & 92 & 87 & 61 & 67 & 62 & 38 & 62 & 67\\
\embed{} & 90 & 73 & 93 & 43 & 76 & 62 & 24 & 38 & 67\\
\midrule
\memzero{} & 90 & 17 & 99 & 46 & 78 & 56 & 22 & 33 & 39\\
\cognee{} & 93 & 84 & 95 & 50 & 70 & 57 & 30 & 52 & 65\\
\memoryos{} & 80 & 76 & 75 & 27 & 85 & 65 & 15 & 35 & 50\\
\lightmem{} & 56 & 8 & 61 & 26 & 61 & 83 & 9 & 9 & 48\\ 
\bottomrule
\end{tabular}
\end{table}

\stitle{Rules and derived states are hard to return.}
The targeted columns of Table~\ref{tab:q2-classes} show what each SUT can
return when asked directly. Facts and completeness facts are mostly returned.
Rules are not. \memzero answers \DiagRuleMemZero{}\% of rule probes and
\lightmem{} \DiagRuleLightMem{}\%, so their derivations lose the rules they
apply. Derived states are returned for only \DiagDerivedBand{}\% of probes,
and \lightmem{} returns only \DiagFactLightMem{}\% of facts.

\stitle{Most failures are ungathered.}
Every wrong answer has at least one frontier node. For memory systems,
\DiagUngatheredMemoryBand{}\% of wrong answers contain an ungathered state,
which the SUT returns when asked for it directly but which the request for the
state that needs it does not gather. \lightmem{} differs, as
\DiagUnretrievableLightMem{}\% of its wrong answers contain an unretrievable
fact, which it cannot return even when asked directly. Misderived states occur
most often for \bmtf{}, which gathers its evidence most completely.

\begin{figure}[t]
\centering
\includegraphics[width=0.96\columnwidth]{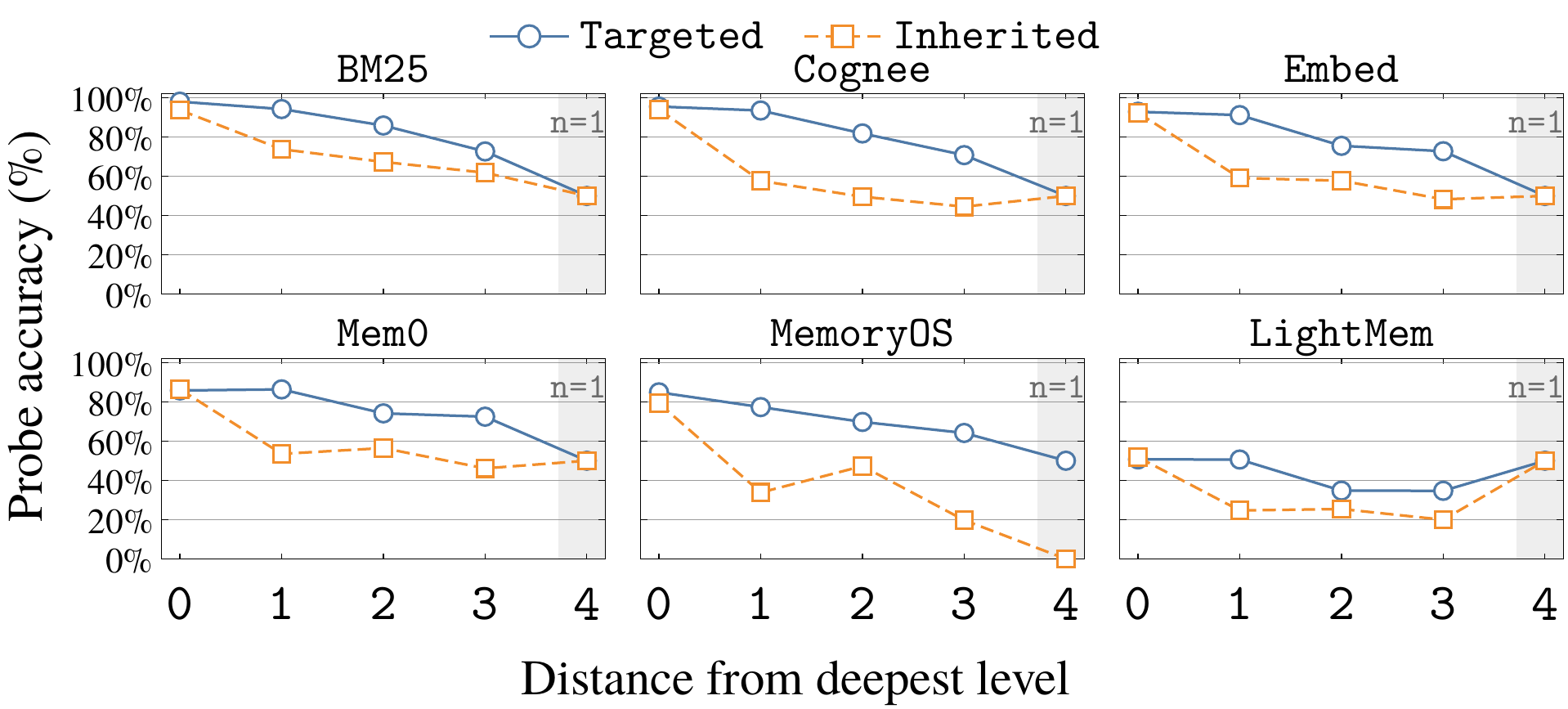}\par
{\small\textbf{(a) Targeted versus inherited context}}\par
\includegraphics[width=\columnwidth,trim={0 10pt 0 8pt},clip]{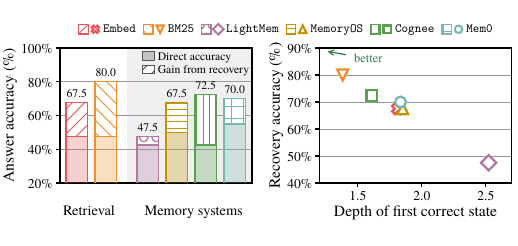}\par
\makebox[0.602\columnwidth]{\small\textbf{(b) Self-evidence recovery}}%
\makebox[0.398\columnwidth]{\small\textbf{(c) First correct state}}
\Description{Accuracy of intermediate states in the targeted and inherited
contexts by distance from the deepest level, accuracy before and after
self-evidence recovery, and the relation between the depth of the first
correct state and the accuracy after recovery.}
\vspace{\capabove}
\caption{Failure diagnosis of the six SUTs on the diagnostic queries:
contexts, recovery, and the first correct state.}
\label{fig:q2-diagnostic-results}
\vspace{\capbelow}
\end{figure}

\stitle{Requests for derived states miss their inputs.}
Figure~\ref{fig:q2-diagnostic-results}(a) compares the targeted and inherited
contexts on every edge of \(\mathcal D_q\), by how many levels the child lies
above the deepest level of \(\mathcal D_q\). The shaded distance holds a single
query. The two contexts agree on the deepest level, which holds facts read
from events. One level above, they separate. The targeted context
answers \DiagTargetedBand{}\% of these states for \bmtf{}, \cognee, and
\memzero, whereas the inherited context answers only
\DiagInheritedBand{}\%. Each input can be retrieved alone, but the request
for the derived state does not bring all of them together, as
Property~\ref{prop:support} predicts.

\stitle{The recoveries confirm the classes.}
Self-evidence recovery raises accuracy by up to \ReplayMemoryMaxGain{} points
(Figure~\ref{fig:q2-diagnostic-results}(b)). It repairs
\DiagSelfRecoveryMemoryBand{}\% of the wrong answers of \memzero, \cognee, and
\memoryos{}, whose failures are mostly ungathered, but only
\DiagSelfRecoveryLightMem{}\% for \lightmem{}. Gold-evidence recovery repairs
\DiagGoldRecoveryMemoryBand{}\% of memory-system failures and adds
\DiagGoldGainLightMem{} points for \lightmem{}, so its wrong evidence is
load-bearing. The failures that remain lie in how states are combined.

\stitle{An early correct state does not ensure recovery.}
Figure~\ref{fig:q2-diagnostic-results}(c) relates how close to the root the
first correct state of a failed query lies to the accuracy after self-evidence
recovery. \bmtf{} finds its first correct state \FirstCorrectDepthBM{} levels below
the root on average and recovers to \ReplayBMPost{}\%, whereas \lightmem{} finds
it \FirstCorrectDepthLightMem{} levels below and recovers to
\ReplayLightMemPost{}\%. An SUT whose correct states lie closer to the root
leaves fewer levels to rebuild. Recovery still needs correct states on every
branch, so one correct state is not enough.

\resultfinding{2}{Most failures of memory systems are ungathered. The SUT can
return the missing state when asked for it directly, but the request for the
state that needs it does not gather it, and self-evidence recovery raises
their accuracy by up to \ReplayMemoryMaxGain{} points. \lightmem{} instead loses
evidence at ingestion, so only gold evidence repairs a substantial share of
its failures.}

\subsection{Q3: Derived-State Maintenance}\label{sec:results-maintenance}

\begin{figure}[t]
\centering
\includegraphics[width=\columnwidth]{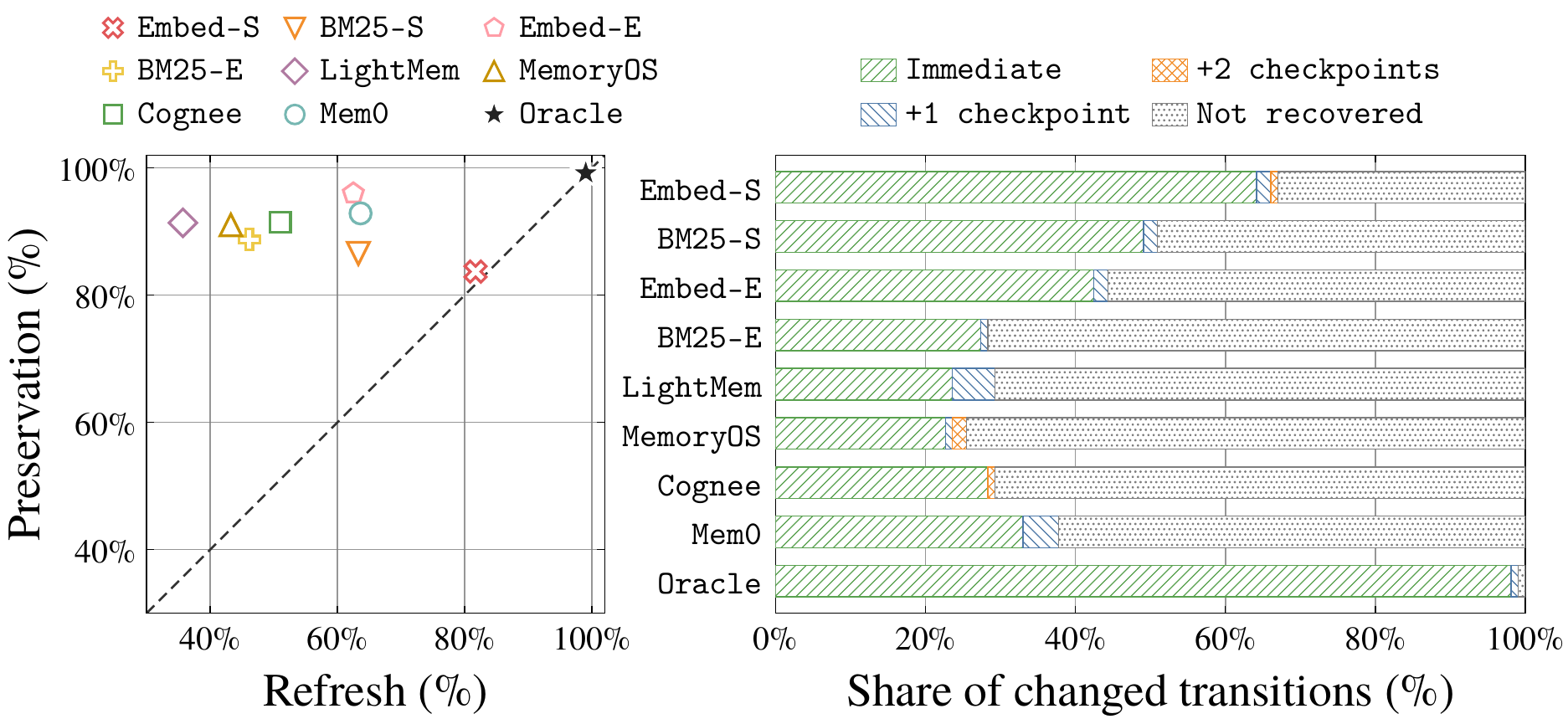}\par
{\small\textbf{(a) Refresh, preservation, and recovery delay}}\par\smallskip
\includegraphics[width=\columnwidth]{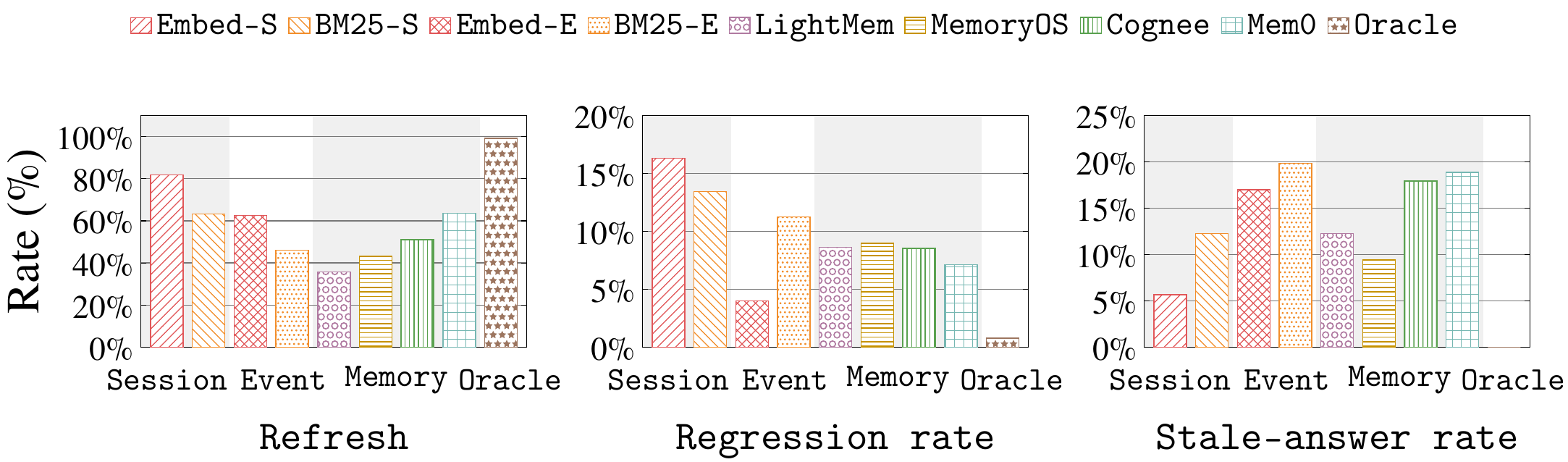}\par
{\small\textbf{(b) Session- and event-level retrieval, memory, and Oracle}}
\Description{Refresh and preservation rates, recovery delay, and the rates
of refresh, regression, and stale answers for session- and event-level
retrieval and the memory systems.}
\vspace{\capabove}
\caption{Maintenance across consecutive checkpoints.}
\label{fig:q3-transition-reliability}
\vspace{\capbelow}
\end{figure}

\stitle{Refresh lags preservation.}
Figure~\ref{fig:q3-transition-reliability}(a) reports both rates of
Equation~\ref{eq:maintenance-rates}. Memory systems preserve
\PreservationBand{}\% of unaffected answers but refresh only \RefreshBand{}\%
of changed ones. \lightmem{} refreshes least and \memzero most.

\stitle{Implicit changes are refreshed least.}
Of the \NumChangedPairs{} changed transitions, \NumExplicitChanges{} are
explicit and \NumImplicitChanges{} are implicit. Memory systems refresh
\RefreshExplicitMemory{}\% of explicit changes but only
\RefreshImplicitMemory{}\% of implicit ones (Table~\ref{tab:q3-change-kinds}).
Raw retrieval refreshes \RefreshExplicitRetrieval{}\% and
\RefreshImplicitRetrieval{}\%. Memory systems thus update what a new record
states but rarely propagate a change through dependencies, which is the lag
that Section~\ref{sec:strategies} attributes to materialization. Raw
retrieval derives at query time and loses less. Stale answers point the same
way. Memory systems repeat the previous answer mostly after
implicit changes, whereas raw retrieval goes stale mostly after explicit ones,
because the superseded record still matches the query.

\begin{table}[t]
\caption{Refresh and stale answers by kind of change (\%).}
\label{tab:q3-change-kinds}
\footnotesize\centering
\setlength{\tabcolsep}{5pt}
\renewcommand{\arraystretch}{0.9}
\begin{tabular}{@{}lrrrr@{}}
\toprule
 & \multicolumn{2}{c}{Explicit} & \multicolumn{2}{c}{Implicit}\\
\cmidrule(lr){2-3}\cmidrule(l){4-5}
SUT & Refresh & Stale & Refresh & Stale\\
\midrule
\bmtf{} & 75 & 28 & 61 & 9\\
\bmtfE{} & 78 & 22 & 37 & 19\\
\embed{} & 83 & 6 & 81 & 6\\
\embedE{} & 100 & 6 & 55 & 19\\
\midrule
\memzero{} & 90 & 17 & 56 & 19\\
\cognee{} & 80 & 11 & 44 & 19\\
\memoryos{} & 78 & 6 & 32 & 10\\
\lightmem{} & 80 & 0 & 26 & 15\\
\midrule
\oraclemode{} & 100 & 0 & 99 & 0\\ 
\bottomrule
\end{tabular}
\end{table}

\stitle{Missed updates persist.}
Figure~\ref{fig:q3-transition-reliability}(b) adds the regression and
stale-answer rates. Memory systems repeat the previous gold answer in
\MemoryStaleBand{}\% of changed transitions. Recovery delay shows that, by the last checkpoint,
\MemoryUnrecoveredBand{} of the \NumChangedPairs{} changes remain unrecovered
for memory systems, against \OracleUnrecovered{} for \oraclemode.

\begin{figure}[t]
\centering
\includegraphics[width=\columnwidth]{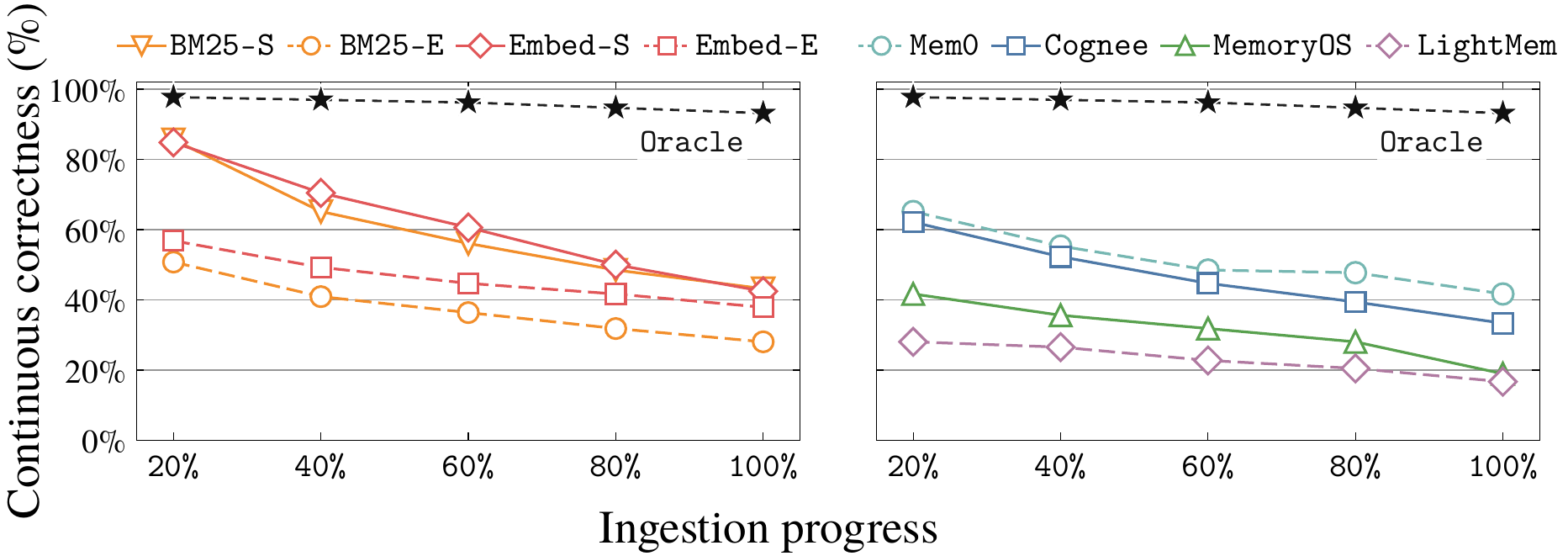}
\Description{The share of tracked queries answered correctly at every
checkpoint up to the displayed one.}
\vspace{\capabove}
\caption{Continuous correctness of the tracked queries.}
\label{fig:q3-checkpoint-prefix-all132}
\vspace{\capbelow}
\end{figure}

\stitle{Few answers stay correct throughout.}
Figure~\ref{fig:q3-checkpoint-prefix-all132} reports continuous correctness.
Memory systems fall from \TrajectoryMemoryFirstBand{}\% at the first
checkpoint to \TrajectoryMemoryFinalBand{}\% at the last, raw retrieval ends
near \TrajectoryBMFinal{}\%, and \oraclemode keeps \TrajectoryOracleFinal{}\%.

\resultfinding{3}{Memory systems preserve unchanged answers well but refresh
only \RefreshImplicitMemory{}\% of implicit changes, half their rate on
explicit ones, so few tracked queries stay correct at every checkpoint.}

\subsection{Ingestion and Query-Time Cost}\label{sec:results-cost}

\begin{figure}[t]
\centering
\includegraphics[width=\columnwidth]{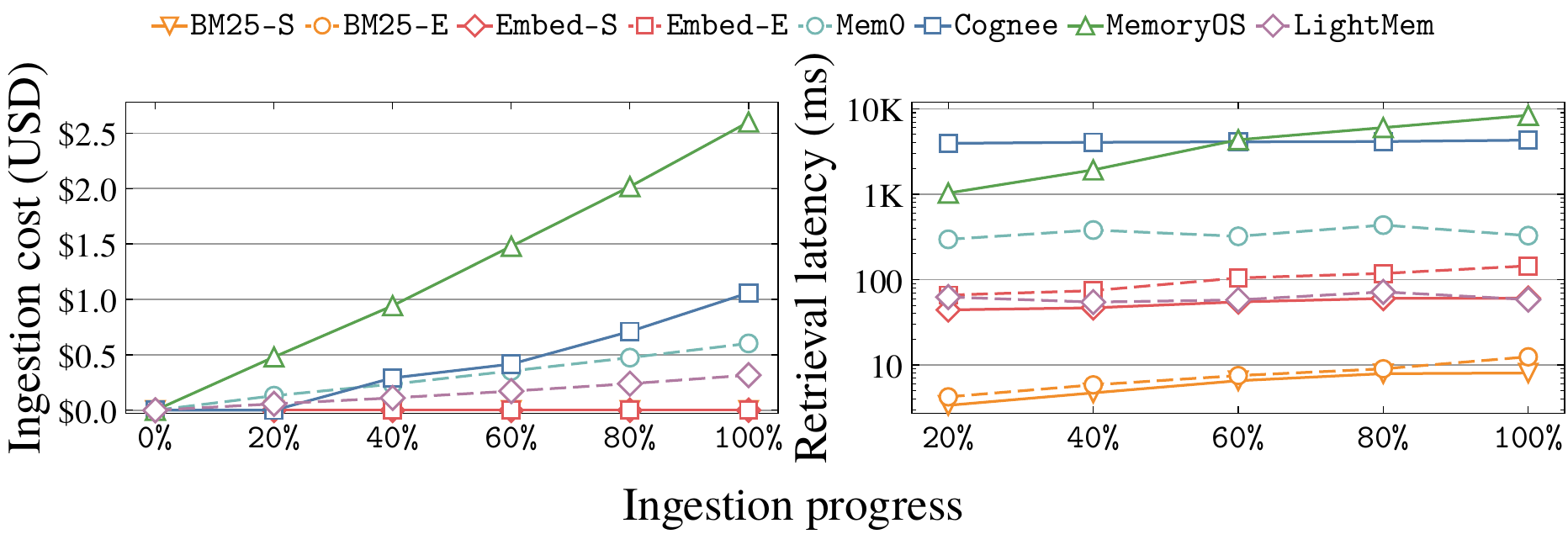}
\Description{Cumulative ingestion cost and median retrieval latency as the
histories grow.}
\vspace{\capabove}
\caption{Ingestion cost and retrieval latency of the SUTs as the histories
grow.}
\label{fig:q3-checkpoint-cost}
\vspace{\capbelow}
\end{figure}

\stitle{Memory systems shift cost to ingestion.}
Figure~\ref{fig:q3-checkpoint-cost} follows the ingestion checkpoints. Memory
systems pay for model calls at ingestion, up to \$\CheckpointMemoryOSCost{}
for \memoryos{}, whereas \bmtf{} and \embed{} make none. Their median
retrieval latency reaches seconds for \cognee and \memoryos{}, against
\CheckpointBMLatency{} ms for \bmtf{}.

\begin{figure}[t]
\centering
\includegraphics[width=\columnwidth]{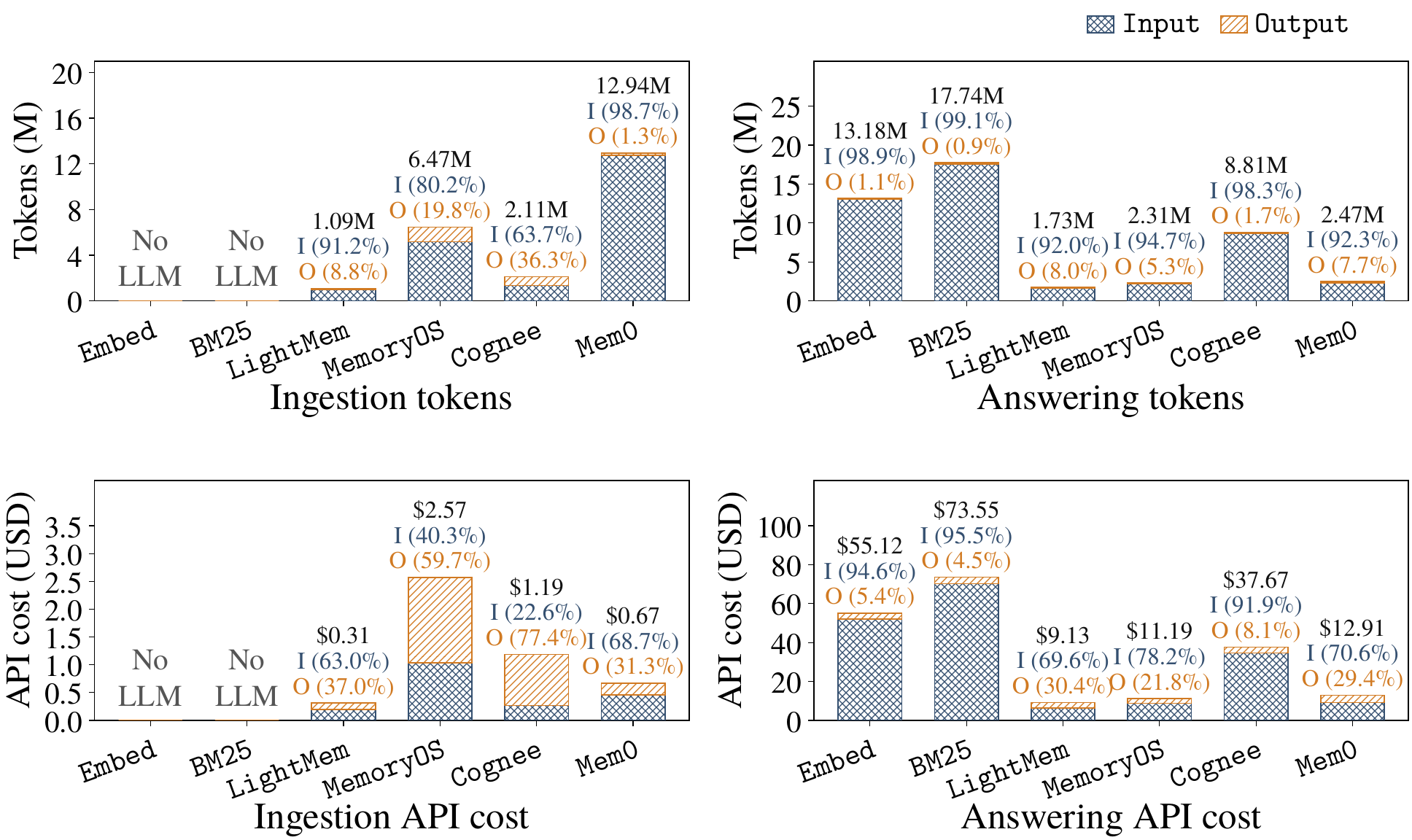}
\Description{Input and output tokens and estimated API cost, separated into
ingestion and answering over all queries.}
\caption{Token and API cost of ingestion and answering.}
\label{fig:q3-token-cost}
\end{figure}

\stitle{Ingestion work does not buy reliability.}
Over all queries (Figure~\ref{fig:q3-token-cost}), memory systems spend
\$\EndMemoryIngestCostBand{} at ingestion and \$\EndMemoryReadCostBand{} on
answering, against \$\EndReadBMCost{} for answering with \bmtf{}, because
they pass the answer model less text. This saving is
real, but the work done at ingestion neither closes the accuracy gap of Q1 nor
refreshes implicit changes. Which placement pays off depends on how often the
history changes and how often its states are reused.

\section{Discussion}\label{sec:lessons}

\subsection{Lessons for Memory Systems}

Read together, the results of Q1 to Q3 and the cost measurements yield five
lessons for the design of memory systems.

\stitle{L1: The bottleneck is gathering, not storage.}
Asked directly, \bmtf, \cognee, and \memzero return \DiagTargetedBand\% of
the states one level above the evidence, but only \DiagInheritedBand\% within
the context of the request that needs them, and most failures are ungathered.
A memory should therefore split a request for a derived state along its
derivation into requests for its inputs. Self-evidence recovery, which does
exactly this, raises accuracy by up to \ReplayMemoryMaxGain{} points.

\stitle{L2: Rules must be stored as first-class objects.}
Asked directly, \memzero returns \DiagRuleMemZero\% of rules and \lightmem
\DiagRuleLightMem\%. Fact-centric extraction summarizes policies and
exceptions away, although every derived state applies them. Rules should be
stored with their versions, like facts.

\stitle{L3: Once gathering works, composition becomes the bottleneck.}
\bmtf gathers evidence most completely, yet \DiagMisderivedBM\% of its wrong
answers contain a misderived state, and accuracy falls with derivation width
and on set and aggregate queries. Better retrieval alone therefore has
diminishing returns. Sets, aggregates, and completeness checks call for
explicit operators, and completeness must be recorded so that a system can
tell an empty answer from an unknown one.

\stitle{L4: Materialization without dependency tracking makes state stale.}
Memory systems refresh \RefreshExplicitMemory\% of explicit changes but
\RefreshImplicitMemory\% of implicit ones, whereas raw retrieval refreshes
\RefreshExplicitRetrieval\% and \RefreshImplicitRetrieval\%. Their stale
answers also arise on opposite kinds of change. A memory system must track
what each stored state depends on and recompute it when an input changes, as
incremental view maintenance does~\cite{gupta1993dred,motik2015bf}. Raw
retrieval instead needs version resolution to suppress superseded records.

\stitle{L5: Ingestion should build structure, not summaries.}
Memory systems pay at ingestion and pass the answer model less text, yet they
trail raw retrieval in accuracy and in refresh. Work at ingestion pays off
only when it records what derivations consume, namely versions, rules,
dependencies, and completeness.

\subsection{Scope}

Executable scenarios make every answer and its derivation checkable. They
cover version resolution, rules, dependencies, sets, aggregation, and
completeness, but not subjective preferences or open-ended planning. The
shared answer model isolates what each SUT supplies, and the recoveries use
the decomposition that the benchmark provides.

\section{Related Work}\label{sec:related}

\begin{table}[t]
\caption{Evaluation properties of memory benchmarks, as reported in the cited
papers.}
\label{tab:comparison}
\centering
\footnotesize
\setlength{\tabcolsep}{2.4pt}
\newcommand{\hd}[1]{\shortstack{#1}}
\begin{tabular}{@{}lccccc@{}}
\toprule
 & \multicolumn{2}{c}{Scoring} & \multicolumn{3}{c}{Tests}\\
\cmidrule(lr){2-3}\cmidrule(l){4-6}
Benchmark & \hd{Answer} & \hd{Evidence} & \hd{Version\\updates} & \hd{State\\maintenance} & \hd{Derived-state\\probes}\\
\midrule
\locomo~\cite{maharana2024locomo} & exact & retrieval & -- & -- & --\\
\longmemeval~\cite{wu2025longmemeval} & judge & retrieval & \yes & -- & --\\
MemoryAgentBench~\cite{hu2025memoryagentbench} & mixed & -- & \yes & -- & --\\
MemTrace~\cite{long2026memtrace} & judge & retrieval & \yes & \yes & --\\
Ground Truth First~\cite{spencer2026groundtruth} & judge & facts & \yes & \yes & --\\
StateMemBench~\cite{fan2026statemembench} & judge & -- & \yes & -- & --\\
\midrule
\sys{} (this paper) & exact & citations & \yes & \yes & \yes\\
\bottomrule
\end{tabular}
\end{table}

\stitle{Agent memory systems.}
Memory systems differ in how they organize a
history~\cite{zhou2026memoryllmera}. \memgpt~\cite{packer2023memgpt} manages
a hierarchy of memory resources, \memzero~\cite{chhikara2025mem0} extracts and
updates salient facts, MemoryLLM~\cite{wang2024memoryllm} keeps memory in
model parameters, \memoryos~\cite{kang2025memoryos} and
\lightmem~\cite{fang2026lightmem} stage memory across levels, and
\cognee~\cite{markovic2025cognee}, \zep~\cite{rasmussen2025zep}, and
AriGraph~\cite{anokhin2025arigraph} link stored information in graphs.
StateMemBench~\cite{fan2026statemembench} and STALE~\cite{chao2026stale}
make supersession, dependencies, and invalidation explicit. \sys prescribes none of
these representations. It evaluates their observable effect, whether the
requested state is correct, whether its inputs can be recovered, and whether
later changes reach it.

\stitle{Memory benchmarks.}
LoCoMo~\cite{maharana2024locomo}, LoCoMo-Plus~\cite{li2026locomoplus}, and
LongMemEval~\cite{wu2025longmemeval} test recall, temporal and multi-session
reasoning, knowledge updates, and abstention over long conversations, and
MemoryAgentBench~\cite{hu2025memoryagentbench} adds
incremental evaluation of retrieval, learning, and forgetting.
MemBench~\cite{tan2025membench}, MEMTRACK~\cite{memtrack2025}, LongMemEval-V2~\cite{wu2026longmemevalv2}, MemoryArena~\cite{memoryarena2026}, AMemGym~\cite{jiayang2026amemgym}, and AMA-Bench~\cite{zhao2026amabench} move
these tests into agentic tasks. Recent benchmarks target specific
failures. StateMemBench~\cite{fan2026statemembench} separates current from superseded
answers, MemTrace~\cite{long2026memtrace} audits how evidence is reached and
used, StreamMemBench~\cite{liu2026streammembench} follows the reuse of
feedback, InMind~\cite{li2026inmind} separates missing facts from facts that
are stored but not surfaced, and Ground Truth
First~\cite{spencer2026groundtruth} checks memory against a script generated
before the text. Table~\ref{tab:comparison} compares their
evaluations. Answers are scored by exact comparison, by a model judge, or by a
mix of both. Evidence names what is checked besides the answer, namely
retrieved sources, stored facts, or cited events against witnesses. The tests
mark whether a benchmark evaluates revised versions of facts, maintenance of
answers as the history grows, and probes of intermediate derived states. \sys
adds a reference derivation for every query, which makes intermediate states
testable and is recomputed at every checkpoint.

\stitle{Provenance and view maintenance.}
Database provenance explains how query results, including aggregates, depend on
input records~\cite{green2007provenance,amsterdamer2011aggregate}, and why and
why-not provenance explain returned and missing
results~\cite{buneman2001why,lee2019pug}. Incremental view maintenance updates
derived results when base data changes~\cite{gupta1993dred,motik2015bf},
including higher-order delta processing and stream
computation~\cite{koch2014dbtoaster,budiu2023dbsp}. \sys builds its witnesses on
minimal why-provenance and recomputes gold as the history grows. It applies
these ideas to memory over natural-language records, where the SUT must first
recover the facts that a database would be given.

\section{Conclusion}\label{sec:conclusion}

Agent memory must supply derived state, which no record states and which
changes as the history evolves. \sys makes this requirement measurable.
Symbolic scenarios give every query an exact answer, a gold derivation DAG,
and its supporting events, so the benchmark can check final answers, locate
the intermediate state where a derivation fails, and follow answers across
updates. Our evaluation shows that current memory systems trail raw retrieval
on derived state, that most of their failures are intermediate states they
can return but do not retrieve for the question that needs them, and that
they rarely propagate implicit changes. Reliable agent memory therefore needs
dependency-aware gathering, versioned rules, and incremental maintenance of
derived state, much as a database maintains views over changing tables.

\bibliographystyle{ACM-Reference-Format}
\bibliography{main}

\appendix
\section{Operators}\label{sec:operator-list}

Table~\ref{tab:operator-list} lists every operator of the released queries.
A query can use several operators, so the counts sum to more than
\NumQueriesTotal. Each operator appears under the family that uses it most.
The shared operators, which mark completeness and plain lookups, occur in
every family.

\begin{table}[!htbp]
\caption{Operators by family, with the number of queries that use each.}
\label{tab:operator-list}
\footnotesize\centering
\setlength{\tabcolsep}{4pt}
\renewcommand{\arraystretch}{1.25}
\begin{tabular}{@{}l>{\raggedright\arraybackslash}p{0.8\columnwidth}@{}}
\toprule
Family & Operators (queries)\\
\midrule
\kw{temporal} & \kw{update}\,(71), \kw{refund}\,(23), \kw{correction}\,(19), \kw{rebooking}\,(12), \kw{expiration}\,(8)\\
\kw{semantic} & \kw{upward\_abstraction}\,(160), \kw{cancellation}\,(61), \kw{scope}\,(37), \kw{containment}\,(32), \kw{personal\_exclusion}\,(28), \kw{custom\_category}\,(20), \kw{derived\_category}\,(8), \kw{type\_rule\_propagation}\,(8), \kw{role\_chain}\,(6)\\
\kw{closure} & \kw{prerequisite\_chain}\,(112), \kw{dependency\_chain}\,(23), \kw{role\_inheritance}\,(11), \kw{renewal}\,(9)\\
\kw{rules} & \kw{rule\_update}\,(153), \kw{exception}\,(138), \kw{custom\_taxonomy}\,(86), \kw{threshold}\,(78), \kw{rule\_selection}\,(56), \kw{conversion}\,(8), \kw{rule\_override}\,(7)\\
\kw{set\_quant} & \kw{aggregation}\,(374), \kw{set\_construction}\,(169), \kw{exclusions}\,(162), \kw{comparison}\,(42), \kw{stale\_exclusion}\,(33), \kw{set\_difference}\,(32), \kw{gating}\,(14), \kw{conjunction}\,(4)\\
\midrule
shared & \kw{closed\_world}\,(186), \kw{lookup}\,(160), \kw{incomplete}\,(141), \kw{negative}\,(79)\\ 
\bottomrule
\end{tabular}
\end{table}

\end{document}